\documentclass[iop]{emulateapj}
\usepackage{color}
\usepackage[hyperfootnotes]{hyperref}

\newcommand{\idrop}{$i_{814}$-dropout}

\newcommand{\ydrop}{$Y_{105}$-dropout}

\newcommand{\yFilter}{$Y_{105}$}
\newcommand{\jFilter}{$J_{125}$}
\newcommand{\jhFilter}{$JH_{140}$}

\newcommand{\hFilter}{$H_{160}$}

\shorttitle{The sizes of $z\sim6-8$ lensed galaxies from the HFF}
\shortauthors{Kawamata et al.}

\begin{document}

\title{The sizes of $z\sim6-8$ lensed galaxies from the Hubble Frontier Fields Abell 2744 data}

\author{Ryota~Kawamata\altaffilmark{1},
Masafumi~Ishigaki\altaffilmark{2,3}, 
Kazuhiro~Shimasaku\altaffilmark{1,4},
Masamune~Oguri\altaffilmark{2,4,5}, 
and Masami~Ouchi\altaffilmark{3,5}
}

\altaffiltext{}{Email: kawamata@astron.s.u-tokyo.ac.jp}
\altaffiltext{1}{Department of Astronomy, Graduate School of Science, The University of Tokyo, 7-3-1 Hongo, Bunkyo-ku, Tokyo 113-0033, Japan}
\altaffiltext{2}{Department of Physics, Graduate School of Science, The University of Tokyo, 7-3-1 Hongo, Bunkyo-ku, Tokyo 113-0033, Japan}
\altaffiltext{3}{Institute for Cosmic Ray Research, The University of Tokyo, 5-1-5 Kashiwanoha, Kashiwa, Chiba 277-8582, Japan}
\altaffiltext{4}{Research Center for the Early Universe, The University of Tokyo, 7-3-1 Hongo, Bunkyo-ku, Tokyo 113-0033, Japan}
\altaffiltext{5}{Kavli Institute for the Physics and Mathematics of the Universe (Kavli IPMU, WPI), The University of Tokyo, 5-1-5 Kashiwanoha, Kashiwa, Chiba 277-8583, Japan}

\begin{abstract}
We investigate sizes of $z\sim6-8$ dropout galaxies using 
the complete data of the Abell 2744 cluster and parallel fields 
in the Hubble Frontier Fields program.
By directly fitting light profiles of observed galaxies with lensing-distorted
S\'ersic profiles on the image plane with the \texttt{glafic} software,
we accurately measure intrinsic sizes
of 31 $z\sim6-7$ and eight $z\sim8$ galaxies,
including those as faint as $M_{\mathrm{UV}}\simeq-16.6$.
We find that half-light radii $r_\mathrm{e}$ positively correlates with UV luminosity
at each redshift, although the correlation is not very tight.
Largest ($r_\mathrm{e}>0.8$ kpc) galaxies are mostly red in UV color 
while smallest ($r_\mathrm{e} < 0.08$ kpc) ones tend to be blue. 
We also find that galaxies with multiple cores tend to be brighter.
Combined with previous results at $2.5\lesssim z\lesssim12$,
our result confirms that the average $r_{\mathrm{e}}$ of bright 
($(0.3-1)L^\star_{z=3}$) galaxies
scales as $r_{\mathrm{e}}\propto(1+z)^{-m}$ with $m=1.24\pm0.1$.
We find that the ratio of $r_\mathrm{e}$ to virial radius 
is virtually constant at $3.3\pm0.1\%$ over 
a wide redshift range, 
where the virial radii of hosting dark matter halos are derived 
based on the abundance matching.
This constant ratio is consistent with the disk formation 
model by \citet{mmw98} with $j_\mathrm{d}\sim m_\mathrm{d}$,  
where $j_\mathrm{d}$ and $m_\mathrm{d}$ are the fractions of 
the angular momentum and mass within halos confined in the disks.
A comparison with various types of local galaxies  
indicates that our galaxies are most similar to 
circumnuclear star-forming regions of barred galaxies 
in the sense that a sizable amount of stars are forming 
in a very small area.

\end{abstract}


\section{Introduction}
The size of galaxies and its redshift evolution provide 
fundamental information on the evolution of galactic structure.
Size measurements of high-$z$ galaxies are particularly important 
for understanding the early phase in the formation of galactic disks.

The high resolution of the Hubble Space Telescope (HST) and sensitive 
cameras installed on it have made it possible to measure the size 
of high-$z$ galaxies.
With the GOODS data from the Advanced Camera for Surveys (ACS), 
\citet{ferg04} have investigated galaxy sizes over $z \sim 1-5$  
to find that the average half-light radius, $r_\mathrm{e}$, of bright 
($\sim L^\star_{z=3}$, 
where $L^{*}_\mathrm{z=3}$ is the characteristic UV luminosity 
of $z\sim3$ LBGs derived in \citealp{steidel99}) galaxies decreases with redshift 
approximately as $(1+z)^{-m}$ with $m \simeq 1.5$.
\citet{bouw04} have also found a similar scaling but with $m\simeq 1$ 
for $z \sim 2.5-6.0$ galaxies in the Hubble Ultra Deep Field data. 
These scalings with redshift imply that galaxies at a fixed luminosity 
evolve in dark matter halos of a constant mass (in the case 
of $m=1$) or a constant circular velocity ($m=1.5$)  
on the assumption that their half-light radii scale linearly 
with the virial radii of their hosting dark matter halos, 
although this non-trivial assumption needs verification.

The decreasing trend of galaxy sizes has been found to 
continue up to $z \sim 7-12$, with $1 \lesssim m \lesssim 1.5$, 
by several papers based on deep data including 
those from the HUDF09 \citep{beckwith06}, 
CANDELS \citep{koekemoer11, grogin11},
and HUDF12 \citep{ellis13, koekemoer13} taken with the 
WFC3/IR camera  
\citep[e.g.][]{oesch10, graz12, ono13, holwerda14},
although \citet{curtislake14} have recently found that the 
modal value of the log-normal distribution of half-light radii 
does not significantly evolve over $z\sim4-8$.
In particular, \citet{ono13} have accurately measured sizes
for nine $z\sim7$ and six $z\sim8$ galaxies from 
the HUDF12 using \texttt{GALFIT} \citep{peng02,peng10}, 
to provide the first evidence for continuation of
the decreasing trend beyond $z \simeq 8$. 
They have also found a clear size--luminosity relation at 
$z\sim 7-8$, which implies that the star formation rate ($S\!F\!R$) correlates 
with the star-formation rate surface density ($\Sigma_\mathrm{SFR}$).

However, beyond $z \sim 6$, the existing samples of accurate 
size measurements are not large enough to determine 
the size--luminosity relation and its tightness.
New samples from sky areas other than the current deep fields 
are needed to improve the statistics and reduce the effect 
of cosmic variance.
The larger sample of size measurements also enables 
one to extend the analysis by, for example, examining 
the dependence of the size--luminosity relation on other 
physical quantities such as galaxy colors.

Besides exploiting the instrumental development, 
using gravitational lensing (GL) is also effective to probe 
high-$z$ galaxies. 
One can investigate detailed properties of intrinsically faint, 
high-$z$ galaxies if they are magnified by foreground lens objects.
Recent examples of such studies using cluster strong lensing 
are found in \citet{BN13} and \citet{alavi14}.

The Hubble Frontier Fields (HFF; PI: J. Lotz) is an on-going 
project to observe six high-magnification clusters deeply with the HST.
The main purpose of the HFF is to investigate distant faint galaxies 
in the background of these clusters with help of lensing magnifications. 
The strong GL effect of these clusters combined with the long
exposures enables us to measure sizes for a large number of 
galaxies as faint as those studied in the HUDF12. 
Highly magnified, extremely faint (in intrinsic luminosity) galaxies 
below the HUDF12 detection limit will also be
discovered.

In this paper, we study sizes of $z \sim 6-8$ galaxies using 
the publicly released HFF data of Abell 2744, the first cluster 
for which the observing program is complete.
We measure sizes for 31 galaxies at $z \sim 6-7$ and 
eight at $z \sim 8$.
Combined with the sample from \citet{ono13}, we now have 40 $z \sim 6-7$ 
and 14 $z \sim 8$ galaxies with accurate size measurements. 

The structure of our paper is as follows.
In Section 2, we construct the galaxy sample for this study, and 
measure intrinsic sizes and luminosities by carefully taking account of  
the GL effect including the magnification and distortion, as well as 
correcting for systematic biases inherent in the photometry 
of faint objects.
Results and discussions are given in Section 3.
Section 3.1 examines the size--luminosity relation at $z \sim 6-8$.
In Section 3.2, we discuss size evolution based on our data and 
literature data over a wide redshift range.
The scaling of galaxy size with the size of hosting dark matter 
halos is also derived.
Section 3.3 compares the state of star formation between 
$z\sim 6-8$ galaxies and various types of local galaxies 
in the $S\!F\!R$--$\Sigma_\mathrm{SFR}$ plane.
Conclusions are shown in Section 4.

Throughout this paper, 
we adopt a cosmology with $\Omega_{M} = 0.3, \Omega_{\Lambda} = 0.7, 
H_{0} = 70$ km s$^{-1}$ Mpc$^{-1}$.
Magnitudes are given in the AB system \citep{okeg83}.
Galaxy sizes are measured in the physical scale.
In a companion paper \citep{ishigaki15}, 
we derive the UV luminosity functions of galaxies at $z \sim 5-10$ 
and discuss their implications for cosmic reionization.

\section{Size measurements of $z\sim 6-8$ galaxies}

\subsection{Sample Selection}

We measure sizes for a bright subset of \idrop\ and \ydrop\ 
galaxies in \citet{ishigaki15}, which are constructed from the version 1.0 
release data of Epoch 1 and Epoch 2 
of the Abell 2744 main cluster and parallel fields.
We use the 30 mas pixel$^{-1}$ image mosaics with the standard
correction. For NIR images of the parallel field, the mosaics 
corrected for time-variable background sky emission are used.
See \citet {ishigaki15} for details of the sample construction.
Briefly, $z\sim 6-7$ galaxies, or \idrop s, are selected by: 
\begin{eqnarray*}
i_{814} - Y_{105} &>& 0.8,\\
Y_{105} - J_{125} &<& 0.8,\\
i_{814} - Y_{105} &>& 2(Y_{105} - J_{125}) + 0.6,\\
S/N(Y_{105}) &>& 5.0, 
S/N(J_{125}) > 5.0,\\
S/N(B_{435}) &<& 2.0, 
S/N(V_{606}) < 2.0, 
\end{eqnarray*}
and $z \sim 8$ galaxies, or \ydrop s, are selected by: 
\begin{eqnarray*}
Y_{105} - J_{125} &>& 0.5,\\
J_{125} - H_{160} &<& 0.4,\\
S/N(J_{125}) &>& 3.5, 
S/N(JH_{140}) > 3.5,\\
S/N(B_{435}) &<& 2.0,
S/N(V_{606}) < 2.0,\\ 
S/N(i_{814}) &<& 2.0.
\end{eqnarray*}
As a result, 18 \idrop s and 11 \ydrop s are selected in the main 
cluster field and 17 \idrop s and four \ydrop s in the parallel field.
It is worth noting that most of the \ydrop s in the cluster
field are highly clustered in a small region. If this overdense 
region is a forming cluster, they could have some different 
properties from galaxies in the average field. This overdensity 
has been reported by some previous studies \citep[e.g.][]{zheng14,ishigaki15}.

Galaxy size measurements require high signal-to-noise ratios.
To maximize the number of galaxies for size measurements, 
we coadd the \yFilter, \jFilter, and \jhFilter\ images 
to make a deep UV-continuum image for the size analysis of 
\idrop s, and coadd the \jFilter, \jhFilter, and \hFilter\ images 
for \ydrop s.
In this process, we use $\texttt{SWarp}$ \citep{bertin02} and 
exploit the inverse variance weight images in the release data set, 
considering differences in exposure time and zero point properly.
The $5\sigma$ limiting magnitudes of the coadded images are 
29.27 (\yFilter\ + \jFilter\ + \jhFilter) and 
29.10 (\jFilter\ + \jhFilter\ + \hFilter) for the main cluster field 
and 29.36 (\yFilter\ + \jFilter\ + \jhFilter) and 
29.07 (\jFilter\ + \jhFilter\ + \hFilter) for the parallel field.

We then run $\texttt{SExtractor}$ \citep{bertin96} 
on the coadded images, and select galaxies brighter than 
the $10\sigma$ limiting magnitudes for size measurements.
After removing a small number of objects falling in the region 
of bright intracluster light (ICL) unsuitable for reliable size 
measurement,
we are left with 31 $z\sim 6-7$ galaxies and eight $z\sim 8$ 
galaxies as given in Tables \ref{tab:candidates7} and 
\ref{tab:candidates8}.
One of the $z\sim 6-7$ galaxies is as faint as $M_\mathrm{UV} 
\simeq -16.6$ with a high magnification of 11.4, where 
$M_{\mathrm{UV}}$ is the rest-frame UV 
($\lambda_{\mathrm{rest}}\simeq 1500\ \mathrm{\AA}$) absolute magnitude. 
This is the faintest $z \gtrsim 6$ galaxy whose size has ever 
been measured.
\citet{ishigaki15} have also selected six $z\sim 9$ galaxies, 
but none of them is included in our analysis because they are 
either too faint or too close to a bright star.

\begin{deluxetable*}{lcclrcrclc} 
\tabletypesize{\scriptsize}
\tablecaption{Dropout candidates at $z\sim 6-7$\label{tab:candidates7} 
in the Abell 2744 field.}
\tablewidth{0pt}
\tablehead{
      \colhead{ID} & \colhead{R.A.\tablenotemark{a}} & \colhead{Dec\tablenotemark{a}} & \colhead{$i_{814}-Y_{105}$} & \colhead{$Y_{105}-J_{125}$} & \colhead{$J_{125}$\tablenotemark{b}} & \colhead{$J_{125} - H_{160}$} & \colhead{Magnification\tablenotemark{c,d}} & \colhead{Photo-$z$} & \colhead{Reference\tablenotemark{e}}
}
\startdata
\sidehead{Cluster field}
HFF1C-i1 & $ 3.593804 $ & $ -30.415447 $ &$ > 2.37 $&$ 0.05 \pm 0.07 $&$ 26.10 \pm 0.05 $& $-0.17^{+0.05}_{-0.05}$ & $3.73^{+0.24}_{-0.21}$ &                  $6.6\pm0.8$    & $1$, $3$\\                    
HFF1C-i2 & $ 3.570654 $ & $ -30.414659 $ &$ 1.33 \pm 0.12 $&$ 0.11 \pm 0.05 $&$ 26.21 \pm 0.03 $& $-0.08^{+0.06}_{-0.06}$ & $1.62^{+0.06}_{-0.06}$ &        $6.0\pm0.7$    & $1$, $3$\\                      
HFF1C-i3 & $ 3.606222 $ & $ -30.386644 $ &$ 1.07 \pm 0.09 $&$ 0.09 \pm 0.04 $&$ 26.25 \pm 0.04 $& $0.07^{+0.06}_{-0.06}$  & $1.69^{+0.05}_{-0.05}$ &        $5.8\pm0.7$            & $1$, $3$\\                
HFF1C-i4 & $ 3.606385 $ & $ -30.407282 $ &$ 1.69 \pm 0.21 $&$ 0.00 \pm 0.05 $&$ 26.37 \pm 0.04 $ & $-0.33^{+0.07}_{-0.07}$& $2.25^{+0.12}_{-0.10}$&            $6.3\pm0.7$        & $1$, $3$\\                  
HFF1C-i5 & $ 3.580452 $ & $ -30.405043 $ &$ > 2.10 $&$ 0.17 \pm 0.07 $&$ 26.60 \pm 0.05 $ & $-0.22^{+0.09}_{-0.09}$& $5.64^{+0.40}_{-0.39}$&          $6.8\pm0.8$        & $1$, $3$\\                
HFF1C-i6 & $ 3.597834 $ & $ -30.395961 $ &$ > 1.71 $&$ 0.27 \pm 0.09 $&$ 26.79 \pm 0.07 $ & $-0.27^{+0.10}_{-0.10}$& $2.87^{+0.19}_{-0.19}$ &                        $7.0\pm0.8$              & $1$, $3$\\              
HFF1C-i7 & $ 3.590761 $ & $ -30.379408 $ &$ 0.95 \pm 0.13 $&$ -0.16 \pm 0.09 $&$ 27.06 \pm 0.08 $ & $-0.13^{+0.12}_{-0.12}$ & $1.87^{+0.06}_{-0.05}$&                $5.9\pm0.7$      & \nodata\\                
HFF1C-i9 & $ 3.601072 $ & $ -30.403991 $ &$ 1.11 \pm 0.27 $&$ 0.00 \pm 0.11 $&$ 27.26 \pm 0.09 $  & $-0.19^{+0.15}_{-0.15}$ & $3.56^{+0.26}_{-0.23}$&  $5.9\pm0.7$        &  $1$, $3$\\                
HFF1C-i10 & $ 3.600619 $ & $ -30.410296 $ &$ > 1.43 $&$ -0.06 \pm 0.11 $&$ 27.29 \pm 0.08 $ & $-0.45^{+0.18}_{-0.18}$& $11.43^{+1.60}_{-1.20}$&      $6.4\pm0.7$        &  $3$\\              
HFF1C-i11 & $ 3.603426 $ & $ -30.383219 $ &$ 0.87 \pm 0.16 $&$ -0.13 \pm 0.11 $&$ 27.29 \pm 0.09 $  & $-0.15^{+0.15}_{-0.15}$ & $1.71^{+0.04}_{-0.05}$&        $5.8\pm0.7$          & $3$\\              
HFF1C-i12 & $ 3.603214 $ & $ -30.410350 $ &$ > 1.36 $&$ -0.03 \pm 0.12 $&$ 27.32 \pm 0.09 $ & $0.10^{+0.14}_{-0.14}$& $3.88^{+0.29}_{-0.21}$&        $6.3\pm0.7$      & $1$, $2$, $3$\\                
HFF1C-i13 & $ 3.592944 $ & $ -30.413328 $ &$ > 1.25 $&$ -0.09 \pm 0.20 $&$ 27.35 \pm 0.15 $ & $-0.03^{+0.16}_{-0.16}$& $6.85^{+0.60}_{-0.54}$ &                        $6.1\pm0.7$    & $3$\\                  
HFF1C-i14 & $ 3.585016 $ & $ -30.413084 $ &$ 0.85 \pm 0.19 $&$ -0.23 \pm 0.14 $&$ 27.45 \pm 0.12 $  & $0.14^{+0.16}_{-0.16}$ & $2.94^{+0.18}_{-0.17}$&                $5.7^{+0.7}_{-1.1}$      & \nodata\\      
HFF1C-i15 & $ 3.576889 $ & $ -30.386329 $ &$ > 0.96 $&$ 0.13 \pm 0.18 $&$ 27.45 \pm 0.16 $& $-0.27^{+0.19}_{-0.19}$& $2.77^{+0.15}_{-0.13}$  &                        $6.1^{+0.8}_{-0.7}$            & $3$\\  
HFF1C-i16 & $ 3.609003 $ & $ -30.385283 $ &$ 1.35 \pm 0.33 $&$ -0.07 \pm 0.14 $&$ 27.56 \pm 0.12 $  & $-0.12^{+0.19}_{-0.19}$& $1.59^{+0.04}_{-0.04}$&              $6.1\pm0.7$          & $3$\\            
HFF1C-i17 & $ 3.604563 $ & $ -30.409364 $ &$ > 0.91 $&$ 0.12 \pm 0.16 $&$ 27.62 \pm 0.11 $ & $-0.25^{+0.22}_{-0.22}$& $2.94^{+0.19}_{-0.16}$ &                        $6.1\pm0.8$        &  $3$\\            
\sidehead{Parallel field}
HFF1P-i1& $ 3.474802 $ & $ -30.362578 $ &$ > 1.80 $&$ 0.46 \pm 0.07 $&$ 26.52 \pm 0.05 $ & $-0.20^{+0.07}_{-0.07}$ & $1.04$& $7.3\pm0.8$ &\nodata\\              
HFF1P-i2& $ 3.480642 $ & $ -30.371175 $ &$ 1.76 \pm 0.34 $&$ -0.02 \pm 0.09 $&$ 26.95 \pm 0.07 $ & $-0.45^{+0.11}_{-0.11}$& $1.05$& $6.3\pm0.7$&\nodata\\    
HFF1P-i3& $ 3.487575 $ & $ -30.364380 $ &$ 1.27 \pm 0.33 $&$ 0.32 \pm 0.11 $&$ 27.06 \pm 0.08 $ & $0.08^{+0.10}_{-0.10}$& $1.05$& $5.8\pm0.7$&\nodata\\      
HFF1P-i4& $ 3.488924 $ & $ -30.394630 $ &$ > 1.39 $&$ 0.25 \pm 0.11 $&$ 27.14 \pm 0.08 $ & $0.03^{+0.11}_{-0.11}$ & $1.05$& $6.7\pm0.8$&\nodata\\              
HFF1P-i5& $ 3.482550 $ & $ -30.371559 $ &$ 1.19 \pm 0.29 $&$ 0.17 \pm 0.11 $&$ 27.15 \pm 0.09 $ & $0.23^{+0.10}_{-0.10}$& $ 1.05$& $5.8^{+0.7}_{-1.4}$&\nodata\\    
HFF1P-i6& $ 3.483960 $ & $ -30.397152 $ &$ > 1.57 $&$ 0.00 \pm 0.11 $&$ 27.20 \pm 0.09 $ & $-0.07^{+0.12}_{-0.12}$ & $1.05$& $6.3\pm0.7$&\nodata\\                
HFF1P-i7& $ 3.467582 $ & $ -30.396908 $ &$ > 1.39 $&$ 0.15 \pm 0.12 $&$ 27.23 \pm 0.09 $ & $-0.19^{+0.13}_{-0.13}$& $1.04$& $6.8\pm0.8$&\nodata\\            
HFF1P-i8& $ 3.467097 $ & $ -30.387686 $ &$ 1.28 \pm 0.25 $&$ -0.24 \pm 0.11 $&$ 27.30 \pm 0.10 $ & $-0.10^{+0.13}_{-0.13}$ & $ 1.04$& $6.0\pm0.7$&\nodata\\      
HFF1P-i9& $ 3.489520 $ & $ -30.399528 $ &$ > 1.41 $&$ 0.05 \pm 0.13 $&$ 27.32 \pm 0.10 $ & $-0.26^{+0.14}_{-0.14}$ & $1.05$& $6.6\pm0.8$&\nodata\\              
HFF1P-i10& $ 3.466056 $ & $ -30.394409 $ &$ 1.10 \pm 0.30 $&$ 0.00 \pm 0.14 $&$ 27.43 \pm 0.11 $ & $-0.10^{+0.15}_{-0.15}$& $1.04$& $6.0\pm0.7$&\nodata\\        
HFF1P-i11& $ 3.460587 $ & $ -30.366320 $ &$ 0.92 \pm 0.34 $&$ 0.05 \pm 0.18 $&$ 27.70 \pm 0.14 $ & $-0.09^{+0.18}_{-0.18}$& $1.04$& $5.8^{+0.7}_{-1.1}$&\nodata\\      
HFF1P-i12& $ 3.455844 $ & $ -30.366359 $ &$ 1.03 \pm 0.36 $&$ 0.00 \pm 0.18 $&$ 27.70 \pm 0.14 $ & $0.41^{+0.16}_{-0.16}$& $1.03$& $4.5^{+1.6}_{-3.9}$&\nodata\\      
HFF1P-i13& $ 3.488139 $ & $ -30.367864 $ &$ > 0.91 $&$ 0.12 \pm 0.19 $&$ 27.73 \pm 0.15 $ & $0.01^{+0.18}_{-0.18}$ & $1.06$& $5.8^{+1.0}_{-5.2}$&\nodata\\            
HFF1P-i14& $ 3.486988 $ & $ -30.399579 $ &$ 0.91 \pm 0.33 $&$ -0.07 \pm 0.18 $&$ 27.75 \pm 0.15 $ & $-0.02^{+0.19}_{-0.19}$ & $1.05$& $5.9^{+0.7}_{-1.1}$&\nodata\\    
HFF1P-i16& $ 3.477238 $ & $ -30.385998 $ &$ > 1.02 $&$ -0.22 \pm 0.21 $&$ 27.98 \pm 0.18 $ & $0.03^{+0.22}_{-0.22}$& $1.05$& $6.5\pm0.7$&\nodata            
\enddata
\tablenotetext{a}{Coordinates are in J2000.}
\tablenotetext{b}{Total magnitude.}
\tablenotetext{c}{The magnification errors in the parallel field are less than $1\%$.}
\tablenotetext{d}{Median value of the magnification distribution.}
\tablenotetext{e}{References: (1) \citet{atek14}; (2) \citet{zheng14}; (3) \citet{atek14b}.}
\end{deluxetable*}

\begin{deluxetable*}{lcclrcrclc}
\tabletypesize{\scriptsize}
\tablecaption{Dropout candidates at $z \sim 8$\label{tab:candidates8} 
in the Abell 2744 field.}
\tablewidth{0pt}
\tablehead{
      \colhead{ID} & \colhead{R.A.\tablenotemark{a}} & \colhead{Dec\tablenotemark{a}} & \colhead{$Y_{105}-J_{125}$} & \colhead{$J_{125}-H_{160}$} & \colhead{$JH_{140}$\tablenotemark{b}} & \colhead{$JH_{140}-H_{160}$} & \colhead{Magnification\tablenotemark{c,d}} & \colhead{Photo-$z$} & \colhead{Reference\tablenotemark{e}}
}
\startdata
\sidehead{Cluster field}
HFF1C-Y1 & $ 3.604518 $ & $ -30.380467 $ &$ 1.17 \pm 0.06 $&$ 0.04 \pm 0.04 $&$ 25.91 \pm 0.02 $ & $0.01^{+0.04}_{-0.04}$ & $1.49^{+0.04}_{-0.04}$      &      $8.0\pm 0.9 $      & $1$, $2$, $4$, $5$, $6$\\ 
HFF1C-Y2 & $ 3.603378 $ & $ -30.382254 $ &$ 1.26 \pm 0.14 $&$ 0.09 \pm 0.07 $&$ 26.62 \pm 0.05 $ & $-0.06^{+0.08}_{-0.08}$& $1.61^{+0.05}_{-0.05}$    &        $8.2\pm 0.9 $      & $2$, $4$, $6$\\          
HFF1C-Y3 & $ 3.596091 $ & $ -30.385833 $ &$ 1.30 \pm 0.16 $&$ -0.00 \pm 0.08 $&$ 26.67 \pm 0.05 $ & $-0.07^{+0.08}_{-0.08}$& $2.25^{+0.09}_{-0.09}$&    $8.2\pm 0.9 $      & $2$, $4$, $6$\\          
HFF1C-Y4 & $ 3.606461 $ & $ -30.380996 $ &$ 1.12 \pm 0.13 $&$ -0.09 \pm 0.08 $&$ 26.94 \pm 0.06 $ & $0.08^{+0.10}_{-0.10}$ & $1.49^{+0.04}_{-0.04}$    &        $8.0\pm 0.9 $      & $2$, $4$, $6$\\          
HFF1C-Y5 & $ 3.603859 $ & $ -30.382263 $ &$ 1.92 \pm 0.34 $&$ 0.35 \pm 0.10 $&$ 26.98 \pm 0.07 $ & $0.19^{+0.10}_{-0.10}$ & $1.60^{+0.05}_{-0.05}$    &        $8.4\pm 0.9 $      & $2$, $4$, $6$\\          
HFF1C-Y6 & $ 3.606577 $ & $ -30.380924 $ &$ 1.09 \pm 0.22 $&$ 0.40 \pm 0.12 $&$ 27.15 \pm 0.08 $ & $0.17^{+0.12}_{-0.12}$& $1.48^{+0.04}_{-0.04}$      &        $7.9^{+0.9}_{-6.1}  $      & $2$, $6$\\      
\sidehead{Parallel field}
HFF1P-Y1 & $ 3.474918 $ & $ -30.362542 $ &$ 0.61 \pm 0.11 $&$ 0.04 \pm 0.08 $&$ 26.93 \pm 0.07 $& $0.07^{+0.08}_{-0.08}$ & $1.05$& $7.5^{+0.8}_{-1.5}$& \nodata\\    
HFF1P-Y2 & $ 3.459245 $ & $ -30.367360 $ &$ 0.73 \pm 0.18 $&$ 0.02 \pm 0.13 $&$ 27.44 \pm 0.11 $& $0.13^{+0.13}_{-0.13}$ & $1.04$& $7.6^{+0.8}_{-1.7}$& \nodata  
\enddata
\tablenotetext{a}{Coordinates are in J2000.}
\tablenotetext{b}{Total magnitude.}
\tablenotetext{c}{The magnification errors in the parallel field are less than $1\%$.}
\tablenotetext{d}{Median value of the magnification distribution.}
\tablenotetext{e}{References: (1) \citet{atek14}; (2) \citet{zheng14}; (3) \citet{zheng14} possible candidates; (4) \citet{coe14}; (5) \citet{laporte14}; (6) \citet{atek14b}.}
\end{deluxetable*}

\subsection{Measurements of Intrinsic Sizes and Luminosities}

\begin{deluxetable*}{rcccclcc}
\tablecolumns{8}
\tabletypesize{\scriptsize}
\tablecaption{Fitting results for dropouts at $z\sim 6-7$.
\label{tab:i-results}}
\tablewidth{0pt}
\tablehead{
\colhead{ID\tablenotemark{a}} &
\colhead{$m_\mathrm{UV}^\mathrm{AUTO}$\tablenotemark{b}} &
\colhead{$n$} &
\colhead{$m_\mathrm{UV}$\tablenotemark{c}} &
\colhead{$M_\mathrm{UV}$} &
\colhead{$r_\mathrm{e}$\tablenotemark{d}} &
\colhead{$e$} &
\colhead{Magnification\tablenotemark{e,f}}\\
\colhead{} &
\colhead{[mag]} &
\colhead{} &
\colhead{[mag]} &
\colhead{[mag]} &
\colhead{[kpc]} &
\colhead{} &
\colhead{}
}
\startdata
\sidehead{Cluster field}
HFF1C-i1 & $26.20 \pm{0.01}$ & 1.0 & $27.68 ^{+0.04}_{-0.05}$ & $-19.05 ^{+0.04}_{-0.05}$ & $0.16 ^{+0.02}_{-0.02}$ & 0.31 & $3.73 ^{+0.24}_{-0.21} {}^{+0.65}_{-0.55}$ \\ 
HFF1C-i2 & $26.23 \pm{0.01}$ & 1.0 & $26.79 ^{+0.04}_{-0.05}$ & $-19.93 ^{+0.04}_{-0.05}$ & $0.15 ^{+0.03}_{-0.03}$ & 0.90 & $1.62 ^{+0.06}_{-0.06} {}^{+0.15}_{-0.14}$ \\ 
HFF1C-i3 & $26.34 \pm{0.01}$ & 1.0 & $26.76 ^{+0.04}_{-0.05}$ & $-19.96 ^{+0.04}_{-0.05}$ & $0.41 ^{+0.02}_{-0.02}$ & 0.19 & $1.69 ^{+0.05}_{-0.05} {}^{+0.33}_{-0.18}$ \\ 
HFF1C-i4 & $26.75 \pm{0.02}$ & 1.0 & $27.45 ^{+0.04}_{-0.05}$ & $-19.27 ^{+0.04}_{-0.05}$ & $0.14 ^{+0.02}_{-0.02}$ & 0.37 & $2.25 ^{+0.12}_{-0.10} {}^{+1.01}_{-0.18}$ \\ 
HFF1C-i5 & $26.71 \pm{0.02}$ & 1.0 & $28.71 ^{+0.06}_{-0.06}$ & $-18.01 ^{+0.06}_{-0.06}$ & $0.14 ^{+0.02}_{-0.02}$ & 0.42 & $5.64 ^{+0.40}_{-0.39} {}^{+0.11}_{-0.69}$ \\ 
HFF1C-i6 & $27.06 \pm{0.02}$ & 1.0 & $28.25 ^{+0.08}_{-0.07}$ & $-18.47 ^{+0.08}_{-0.07}$ & $0.09 ^{+0.02}_{-0.03}$ & 0.65 & $2.87 ^{+0.19}_{-0.19} {}^{+0.73}_{-0.09}$ \\ 
$^{\ast}$HFF1C-i7 & $26.40 \pm{0.01}$ & 1.0 & $27.13 ^{+0.09}_{-0.15}$ & $-19.59 ^{+0.09}_{-0.15}$ & $0.73 ^{+0.10}_{-0.07}$ & 0.65 & $1.87 ^{+0.06}_{-0.05} {}^{+0.80}_{-0.07}$ \\ 
$^{\ast}$HFF1C-i9 & $27.14 \pm{0.02}$ & 1.0 & $28.00 ^{+0.10}_{-0.17}$ & $-18.72 ^{+0.10}_{-0.17}$ & $0.55 ^{+0.07}_{-0.06}$ & 0.54 & $3.56 ^{+0.26}_{-0.23} {}^{+0.58}_{-0.43}$ \\ 
HFF1C-i10 & $27.83 \pm{0.04}$ & 1.0 & $30.07 ^{+0.09}_{-0.09}$ & $-16.65 ^{+0.09}_{-0.09}$ & $0.03 ^{+0.01}_{-0.01}$ & 0.85 & $11.4 ^{+1.60}_{-1.20} {}^{+0.06}_{-2.55}$ \\ 
HFF1C-i11 & $27.71 \pm{0.04}$ & 1.0 & $28.00 ^{+0.08}_{-0.07}$ & $-18.72 ^{+0.08}_{-0.07}$ & $0.15 ^{+0.03}_{-0.04}$ & 0.40 & $1.71 ^{+0.04}_{-0.05} {}^{+0.46}_{-0.24}$ \\ 
HFF1C-i12 & $26.44 \pm{0.01}$ & 1.0 & $27.91 ^{+0.11}_{-0.11}$ & $-18.81 ^{+0.11}_{-0.11}$ & $0.57 ^{+0.05}_{-0.06}$ & 0.47 & $3.88 ^{+0.29}_{-0.21} {}^{+5.12}_{-0.26}$ \\ 
HFF1C-i13 & $27.69 \pm{0.04}$ & 1.0 & $29.58 ^{+0.06}_{-0.07}$ & $-17.14 ^{+0.06}_{-0.07}$ & $<0.04$ & 0.78 & $6.85 ^{+0.60}_{-0.54} {}^{+1.65}_{-1.31}$ \\ 
HFF1C-i14 & $26.65 \pm{0.01}$ & 1.0 & $28.53 ^{+0.12}_{-0.14}$ & $-18.19 ^{+0.12}_{-0.14}$ & $0.23 ^{+0.07}_{-0.07}$ & 0.51 & $2.94 ^{+0.18}_{-0.17} {}^{+0.56}_{-0.16}$ \\ 
HFF1C-i15 & $27.61 \pm{0.03}$ & 1.0 & $28.96 ^{+0.09}_{-0.11}$ & $-17.76 ^{+0.09}_{-0.11}$ & $<0.06$ & 0.85 & $2.77 ^{+0.15}_{-0.13} {}^{+4.65}_{-0.00}$ \\ 
HFF1C-i16 & $27.82 \pm{0.04}$ & 1.0 & $28.37 ^{+0.09}_{-0.19}$ & $-18.35 ^{+0.09}_{-0.19}$ & $0.08 ^{+0.04}_{-0.04}$ & 0.90 & $1.59 ^{+0.04}_{-0.04} {}^{+0.61}_{-0.16}$ \\ 
HFF1C-i17 & $27.89 \pm{0.04}$ & 1.0 & $29.19 ^{+0.09}_{-0.11}$ & $-17.53 ^{+0.09}_{-0.11}$ & $<0.08$ & 0.89 & $2.94 ^{+0.19}_{-0.16} {}^{+3.70}_{-0.23}$ \\
\sidehead{Parallel field}
$^{\ast}$HFF1P-i1 & $26.36 \pm{0.01}$ & 1.0 & $26.75 ^{+0.05}_{-0.07}$ & $-19.97 ^{+0.05}_{-0.07}$ & $0.52 ^{+0.05}_{-0.06}$ & 0.68 & 1.04 \\ 
HFF1P-i2 & $27.28 \pm{0.03}$ & 1.0 & $27.20 ^{+0.08}_{-0.07}$ & $-19.52 ^{+0.08}_{-0.07}$ & $0.40 ^{+0.04}_{-0.05}$ & 0.42 & 1.05 \\ 
HFF1P-i3 & $26.87 \pm{0.02}$ & 1.0 & $26.92 ^{+0.10}_{-0.10}$ & $-19.80 ^{+0.10}_{-0.10}$ & $0.83 ^{+0.09}_{-0.07}$ & 0.27 & 1.05 \\ 
HFF1P-i4 & $26.91 \pm{0.02}$ & 1.0 & $27.35 ^{+0.12}_{-0.11}$ & $-19.37 ^{+0.12}_{-0.11}$ & $0.60 ^{+0.07}_{-0.10}$ & 0.55 & 1.05 \\ 
HFF1P-i5 & $27.43 \pm{0.03}$ & 1.0 & $27.19 ^{+0.09}_{-0.09}$ & $-19.53 ^{+0.09}_{-0.09}$ & $0.64 ^{+0.07}_{-0.06}$ & 0.21 & 1.05 \\ 
HFF1P-i6 & $27.04 \pm{0.02}$ & 1.0 & $27.21 ^{+0.11}_{-0.11}$ & $-19.51 ^{+0.11}_{-0.11}$ & $0.60 ^{+0.09}_{-0.11}$ & 0.70 & 1.05 \\ 
HFF1P-i7 & $27.23 \pm{0.02}$ & 1.0 & $27.06 ^{+0.12}_{-0.12}$ & $-19.66 ^{+0.12}_{-0.12}$ & $0.81 ^{+0.08}_{-0.10}$ & 0.50 & 1.04 \\ 
HFF1P-i8 & $27.34 \pm{0.03}$ & 1.0 & $27.26 ^{+0.10}_{-0.10}$ & $-19.46 ^{+0.10}_{-0.10}$ & $0.56 ^{+0.07}_{-0.08}$ & 0.44 & 1.04 \\ 
HFF1P-i9 & $27.37 \pm{0.03}$ & 1.0 & $27.54 ^{+0.11}_{-0.12}$ & $-19.18 ^{+0.11}_{-0.12}$ & $0.41 ^{+0.08}_{-0.09}$ & 0.57 & 1.05 \\ 
HFF1P-i10 & $27.70 \pm{0.04}$ & 1.0 & $27.45 ^{+0.12}_{-0.14}$ & $-19.27 ^{+0.12}_{-0.14}$ & $0.45 ^{+0.10}_{-0.11}$ & 0.74 & 1.04 \\ 
HFF1P-i11 & $26.80 \pm{0.02}$ & 1.0 & $28.35 ^{+0.12}_{-0.10}$ & $-18.37 ^{+0.12}_{-0.10}$ & $0.12 ^{+0.07}_{-0.05}$ & 0.90 & 1.04 \\ 
HFF1P-i12 & $27.62 \pm{0.03}$ & 1.0 & $27.58 ^{+0.23}_{-0.18}$ & $-19.14 ^{+0.23}_{-0.18}$ & $0.80 ^{+0.13}_{-0.18}$ & 0.45 & 1.03 \\ 
HFF1P-i13 & $27.45 \pm{0.03}$ & 1.0 & $27.39 ^{+0.20}_{-0.18}$ & $-19.33 ^{+0.20}_{-0.18}$ & $1.04 ^{+0.18}_{-0.17}$ & 0.31 & 1.06 \\ 
HFF1P-i14 & $27.67 \pm{0.04}$ & 1.0 & $27.22 ^{+0.15}_{-0.16}$ & $-19.50 ^{+0.15}_{-0.16}$ & $1.00 ^{+0.16}_{-0.14}$ & 0.26 & 1.05 \\ 
HFF1P-i16 & $27.83 \pm{0.04}$ & 1.0 & $28.07 ^{+0.17}_{-0.19}$ & $-18.65 ^{+0.17}_{-0.19}$ & $0.34 ^{+0.14}_{-0.13}$ & 0.90 & 1.05
\enddata
\tablenotetext{a}{Asterisks indicate galaxies with multiple cores.}
\tablenotetext{b}{\texttt{MAG\_AUTO} magnitude from \texttt{SExtractor}.}
\tablenotetext{c}{Total apparent magnitude from light profile fitting with \texttt{glafic}.}
\tablenotetext{d}{Circularized effective radius, 
              $r_\mathrm{e}^\mathrm{maj} \sqrt{1-e}$, where 
              $r_\mathrm{e}^\mathrm{maj}$ 
              is the radius along the major axis and 
              $e$ the ellipticity.}
\tablenotetext{e}{Best fitted value of magnification.}
\tablenotetext{f}{The first error shows the error in our mass map, and the second error shows the variation in magnification among the other eight public mass maps.}
\end{deluxetable*}

\begin{deluxetable*}{rccccccc}
\tablecolumns{8}
\tabletypesize{\scriptsize}
\tablecaption{Fitting results for dropouts at $z\sim 8$.
\label{tab:y-results}}
\tablewidth{0pt}
\tablehead{
\colhead{ID\tablenotemark{a}} &
\colhead{$m_\mathrm{UV}^\mathrm{AUTO}$\tablenotemark{b}} &
\colhead{$n$} &
\colhead{$m_\mathrm{UV}$\tablenotemark{c}} &
\colhead{$M_\mathrm{UV}$} &
\colhead{$r_\mathrm{e}$\tablenotemark{d}} &
\colhead{$e$} &
\colhead{Magnification\tablenotemark{e,f}}\\
\colhead{} &
\colhead{[mag]} &
\colhead{} &
\colhead{[mag]} &
\colhead{[mag]} &
\colhead{[kpc]} &
\colhead{} &
\colhead{}
}
\startdata
\sidehead{Cluster field}
$^{\ast}$HFF1C-Y1 & $25.97 \pm{0.02}$ & 1.0 & $26.20 ^{+0.03}_{-0.03}$ & $-20.95 ^{+0.03}_{-0.03}$ & $0.24 ^{+0.02}_{-0.02}$ & 0.70 & $1.49 ^{+0.04}_{-0.04} {}^{+0.56}_{-0.11}$ \\ 
$^{\ast}$HFF1C-Y2 & $25.99 \pm{0.02}$ & 1.0 & $26.48 ^{+0.05}_{-0.13}$ & $-20.67 ^{+0.05}_{-0.13}$ & $0.62 ^{+0.05}_{-0.05}$ & 0.60 & $1.61 ^{+0.05}_{-0.05} {}^{+0.55}_{-0.17}$ \\ 
HFF1C-Y3 & $26.58 \pm{0.03}$ & 1.0 & $27.13 ^{+0.06}_{-0.10}$ & $-20.01 ^{+0.06}_{-0.10}$ & $0.38 ^{+0.03}_{-0.03}$ & 0.20 & $2.25 ^{+0.09}_{-0.09} {}^{+3.06}_{-0.26}$ \\ 
HFF1C-Y4 & $26.66 \pm{0.03}$ & 1.0 & $27.27 ^{+0.09}_{-0.12}$ & $-19.88 ^{+0.09}_{-0.12}$ & $0.27 ^{+0.06}_{-0.05}$ & 0.65 & $1.49 ^{+0.04}_{-0.04} {}^{+0.62}_{-0.11}$ \\ 
$^{\ast}$HFF1C-Y5 & $26.16 \pm{0.02}$ & 1.0 & $26.61 ^{+0.10}_{-0.13}$ & $-20.54 ^{+0.10}_{-0.13}$ & $0.84 ^{+0.08}_{-0.09}$ & 0.46 & $1.60 ^{+0.05}_{-0.05} {}^{+0.54}_{-0.17}$ \\ 
HFF1C-Y6 & $26.46 \pm{0.02}$ & 1.0 & $26.59 ^{+0.15}_{-0.20}$ & $-20.55 ^{+0.15}_{-0.20}$ & $0.92 ^{+0.14}_{-0.10}$ & 0.44 & $1.48 ^{+0.04}_{-0.04} {}^{+0.63}_{-0.11}$ \\
\sidehead{Parallel field}
$^{\ast}$HFF1P-Y1 & $27.15 \pm{0.03}$ & 1.0 & $27.14 ^{+0.09}_{-0.09}$ & $-20.00 ^{+0.09}_{-0.09}$ & $0.24 ^{+0.05}_{-0.07}$ & 0.65 & 1.05 \\ 
HFF1P-Y2 & $27.07 \pm{0.03}$ & 1.0 & $27.48 ^{+0.11}_{-0.14}$ & $-19.66 ^{+0.11}_{-0.14}$ & $0.21 ^{+0.07}_{-0.07}$ & 0.90 & 1.04
\enddata
\tablenotetext{a}{Asterisks indicate galaxies with multiple cores.}
\tablenotetext{b}{\texttt{MAG\_AUTO} magnitude from \texttt{SExtractor}.}
\tablenotetext{c}{Total apparent magnitude from light profile fitting with \texttt{glafic}.}
\tablenotetext{d}{Circularized effective radius, 
              $r_\mathrm{e}^\mathrm{maj} \sqrt{1-e}$, where 
              $r_\mathrm{e}^\mathrm{maj}$ 
              is the radius along the major axis and 
              $e$ the ellipticity.}
\tablenotetext{e}{Best fitted value of magnification.}
\tablenotetext{f}{The first error shows the error in our mass map, and the second error shows the variation in magnification among the other eight public mass maps.}
\end{deluxetable*}

To correct for the GL effect on our galaxies, we use 
the cluster mass map obtained in \citet{ishigaki15},
in which the mass model is constructed with the pubic software $\texttt{glafic}$
\citep{oguri10} using all available multiply lensed objects 
in the literature plus three newly identified objects.
This map is found to be in good agreement with those provided 
by other research teams which are posted on the public HFF website\footnote
{\url{http://www.stsci.edu/hst/campaigns/frontier-fields/Lensing-Models}}.

While $\texttt{GALFIT}$ is widely used for analyzing luminosity 
profiles of faint galaxies, in this paper we employ $\texttt{glafic}$
for size measurements as well, because it enables detailed light
profile fitting similar to $\texttt{GALFIT}$ for lensed, hence distorted,
galaxies.
This package finds the best-fit S\'ersic profile parameters for a given, 
lensed galaxy image by simulating many lensed images with different 
profile parameters taking account of the GL effect at the 
position of the galaxy and fitting with them on the observed image.
That is, distorted S\'ersic profiles are directly compared 
on the image plane.
In this sense, this method is more direct than one in which 
observed (distorted) images are fit with undistorted profiles 
created with $\texttt{GALFIT}$ and then the best-fit radius and 
magnitude are corrected for the GL effect simply by 
dividing by the magnification factor \citep[e.g.][]{laporte14}.

In the course of profile fitting, each simulated image 
is convolved with the PSF when compared with the observed light profile.
The PSF profile for each coadded image is determined by stacking 
four stellar objects in the same image.
In case there are nearby foreground galaxies near the target galaxy,
we carefully mask these nearby galaxies during the profile fitting.

In this paper, we fix the S\'ersic index to $n = 1$, the value 
widely used in previous studies \citep[e.g.][]{ono13}. 
We use the half-light radius
to express galaxy size 
following previous studies. 
We have thus six free parameters to determine: 
positions, half-light radius, flux, ellipticity, and position angle.
The upper limit of the ellipticity is set to $0.9$.
For each galaxy, S\'ersic profile fitting is repeated with different 
initial parameter sets until the best-fit parameters converge.
The output fluxes (half-light radii) are converted into absolute 
magnitudes (radii in physical length) assuming that all the 
$z\sim 6-7$ and $z\sim 8$ galaxies are located at $z=6.1$ and $z=8.0$, 
respectively.

Some galaxies, e.g., CY-1, appear to consist of multiple components 
or contain sub-clumps.
Although we treat these complex galaxies as single objects,
we first fit each system with multiple S\'ersic profiles simultaneously 
in order to obtain reliable sky background 
estimates for the field.
This is because the sky background value obtained by this
multi-component fitting is more stable and robust than 
single profile fitting of these complex systems.
We then perform single profile fitting with the fixed sky 
background value derived above.
Results of light profile fitting for all the dropout galaxies 
are shown in Figures \ref{fig:fitresults7_1}--\ref{fig:fitresults8_parallel} in the Appendix.

There are four very small ($r_{\mathrm e} <0.08$ kpc) galaxies at $z \sim 6-7$:
C-i10, C-i13, C-i15, and C-i17.
However, we find only one (C-i10) to be actually resolved and 
calculate for the remaining three an upper limit to the half-light 
radius, from the examination below:
For each object, we replace the real image with a stellar image 
and conduct a profile fitting in the same manner 
as for the real object considering the GL effect.
This stellar image is taken from the same field 
without any luminosity correction.
We repeat this five times using five different stellar images.
Then, if the average of the five output half-light radii is 
significantly smaller than the half-light radius of the real image, 
we consider this galaxy to be resolved. If not, we consider it 
to be unresolved and adopt the average stellar half-light radius 
for the upper limit.
We also confirm that C-i16 is resolved.

We note that C-i5 and C-i6 are lensed images of the same galaxy.
We adopt the average values of these two images for the parameters 
of this galaxy.

\subsection{Error Estimates}

There are three main sources of errors in our size and magnitude
measurements: Statistical and systematic errors in light profile
fitting, internal statistical errors in lensing magnification and
distortion of our mass map, and external systematic errors in lensing
effects coming from different assumptions made in creating mass
maps. In the HFF project, the last source of errors can be estimated
by comparing lensing properties of eight public mass maps that are
constructed independently \citep{ishigaki15}.
These mass maps enable us to discuss the validity of our mass map 
and to estimate systematic errors caused by differences in the 
method of mass map construction.

\begin{figure}[tbp]
  \centering
      \includegraphics[width=\linewidth]{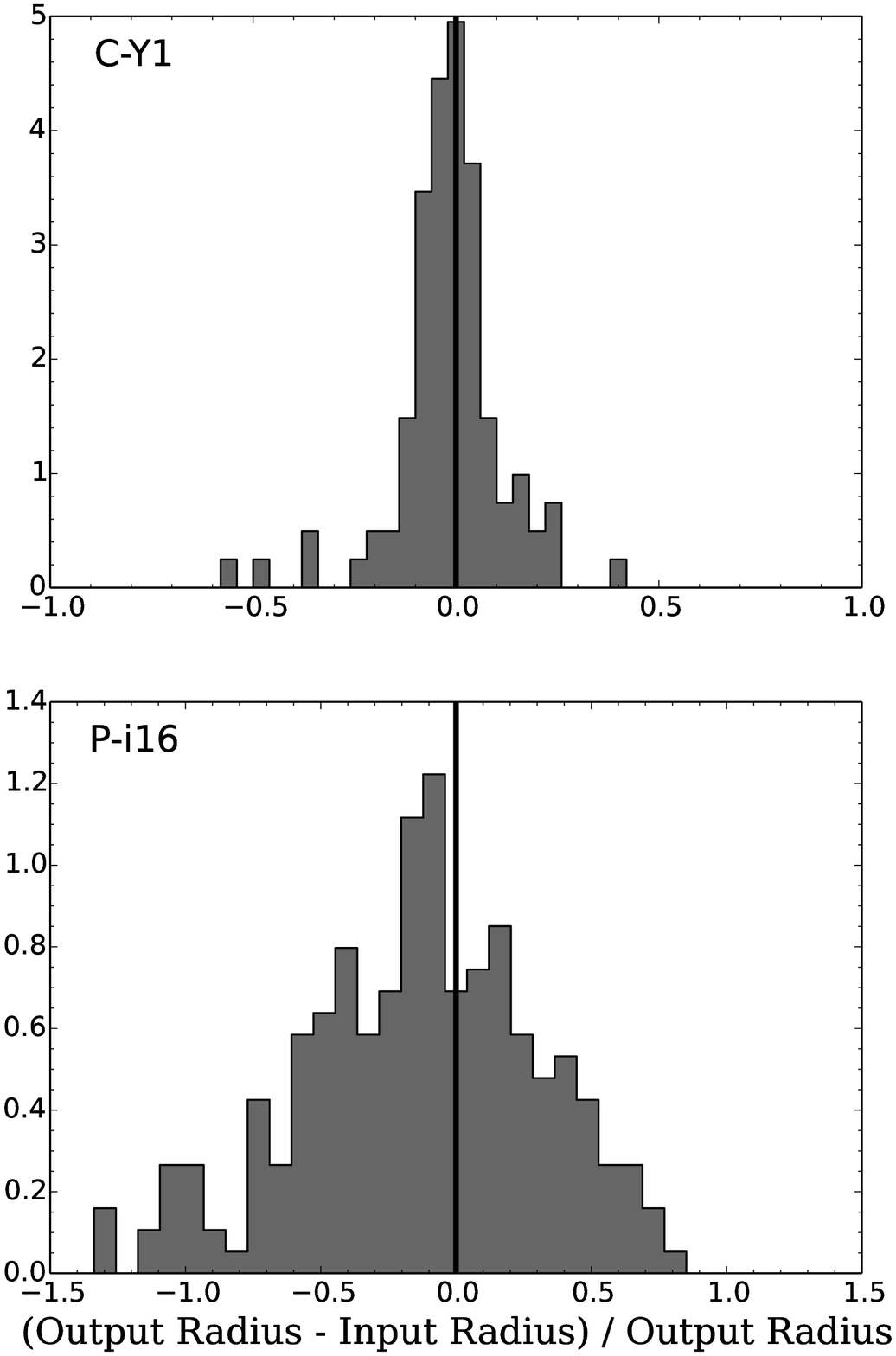}
  \caption{Examples of the simulations to estimate the error 
              in half-light radius.
          The upper and lower panels show
              the normalized distribution of the difference between input 
              and measured radii for C-Y1 and P-i16, respectively.}
  \label{fig:simulation}
\end{figure}

The largest among the three is the error in light profile fitting.
It is well known that the size and brightness of very faint 
galaxies such as those in our sample tend to be biased
depending on their size and magnitude even when they are 
measured from careful profile fitting.
For each galaxy, we correct for these systematic errors 
with the following Monte Carlo approach.
(1) We generate a model galaxy image by randomly changing 
S\'ersic parameter values around the best-fit values, 
and put it randomly near the observed position of the galaxy 
if it is in the main cluster field, where the GL effect 
varies from position to position, or randomly in the entire image 
if it is in the parallel field, where the GL effect is small 
and nearly constant.
In this process, we do not change the GL effect,
which depends on its position, because the purpose here is 
to estimate only photometric errors.
(2) We measure its size and magnitude in exactly the same manner 
as for the observed galaxy.
(3) We repeat (1) and (2) to make a table of 100--300 input and 
measured sizes and magnitudes.
(4) From (3), we extract input half-light radii whose measured 
radii and magnitudes are virtually equal to that of the observed galaxy 
to create their histogram.
The median and the standard deviation of the histogram are 
adopted as the true radius and its error of the galaxy, respectively.
Its true magnitude and error are determined in a similar manner.
As an example, the results of the simulations for C-Y1 
(the most precise case) and P-i16 (the least precise case) are 
shown in Figure \ref{fig:simulation}.

Errors in the mass map affect size and magnitude measurements 
primarily through changes in magnification.
\citet{ishigaki15} have found that the change in magnification 
due to the internal errors in our mass map is negligibly small, 
$\sim 5\%$, for the main cluster field and less than $1\%$ for 
the parallel field, as given in Tables \ref{tab:i-results} and 
\ref{tab:y-results}.

On the other hand, there is a significant variation in 
magnification among the eight public mass maps 
\citep[see Figure 11 of][]{ishigaki15}. We estimate this systematic error 
at the position of each dropout by calculating the standard 
deviation of the eight magnification values excluding the highest
and lowest ones. The errors obtained are shown in Tables \ref{tab:i-results} and \ref{tab:y-results}.
The error bars plotted in Figure \ref{fig:magrad} are a quadratic sum of this 
systematic error and the error in profile fitting.

The best-fit values and errors for all the dropout galaxies are 
summarized in Tables \ref{tab:i-results} and \ref{tab:y-results}.

\section{Results and Discussion}

\subsection{Size--luminosity relation}

\begin{figure}[tbp]
  \centering
      \includegraphics[width=\linewidth]{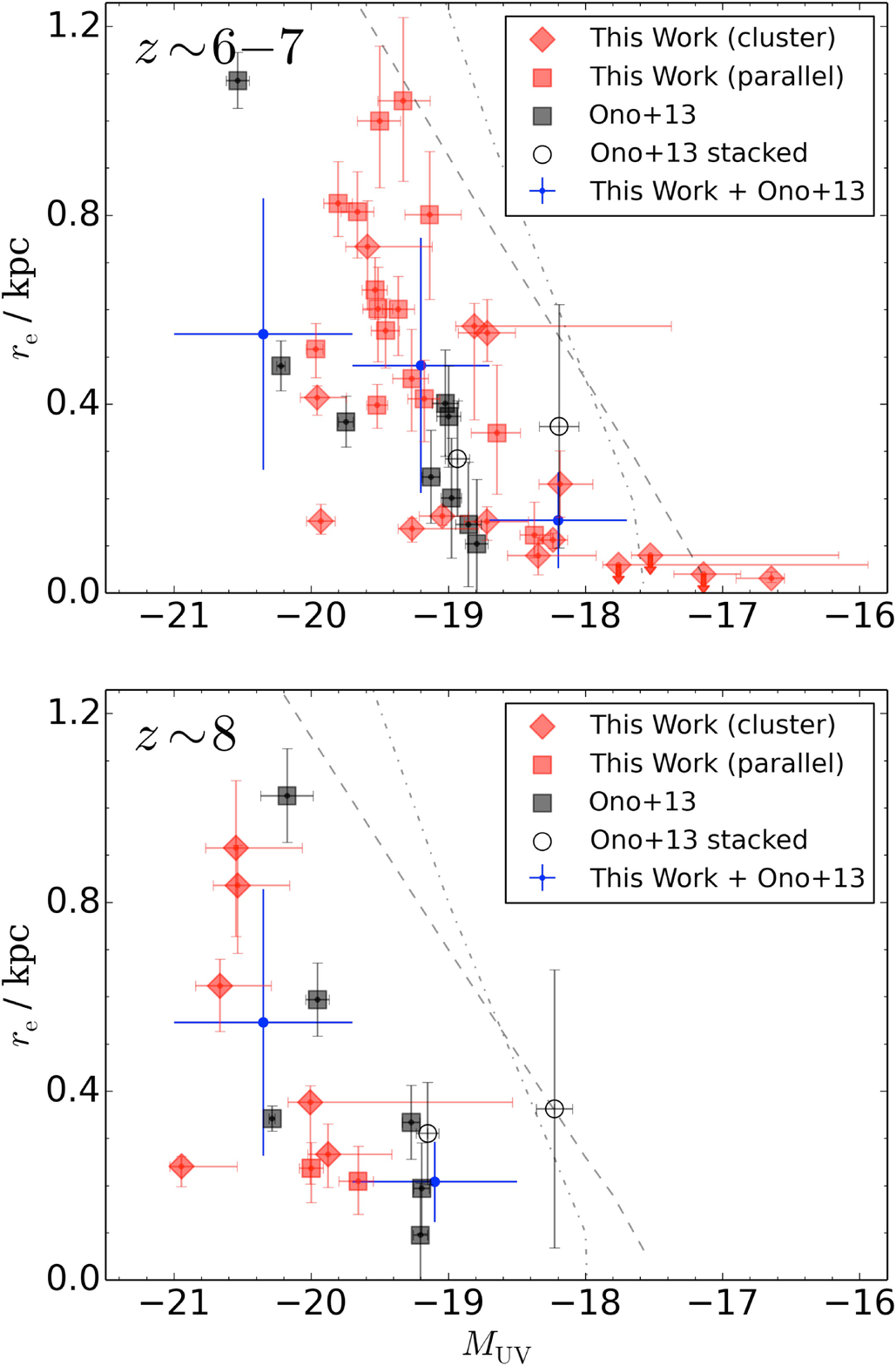}
  \caption{Size--luminosity relation for $z \sim 6-7$ (top) 
    and $z\sim 8$ (bottom) galaxies. 
  The red and gray points represent our galaxies and 
  \citet{ono13}'s, respectively.
  Rhombuses represent galaxies from the cluster field 
  and squares from the parallel and blank fields.
  Arrows represent galaxies 
              whose half-light radius is an upper limit.
  The error bars of our galaxies show conservative errors 
  that include the fitting error and the variation in magnification
  among the eight other mass maps.
  The blue points with bars indicate the average size 
    and the scatter in the given luminosity bin for 
    the merged sample of this study and \citet{ono13}. 
  The dashed and dash-dot lines indicate the 50\% completeness 
    lines for our samples in the cluster and parallel fields, respectively.}
  \label{fig:magrad}
\end{figure}

\begin{figure}[tbp]
  \centering
      \includegraphics[width=\linewidth]{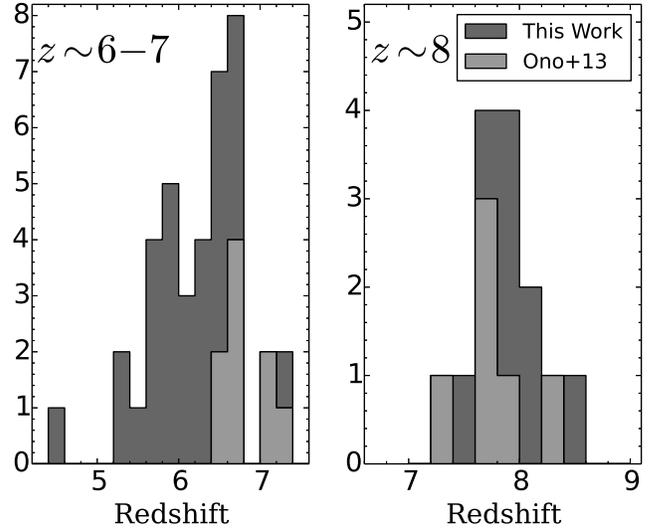}
  \caption{Redshift distribution for $z \sim 6-7$ (left) 
    and $z\sim 8$ (right) galaxies.
    The gray and light gray histograms represent our galaxies and 
  \citet{ono13}'s, respectively.}
  \label{fig:redshift-distribution}
\end{figure}

\begin{figure}[tbp]
  \centering
      \includegraphics[width=\linewidth]{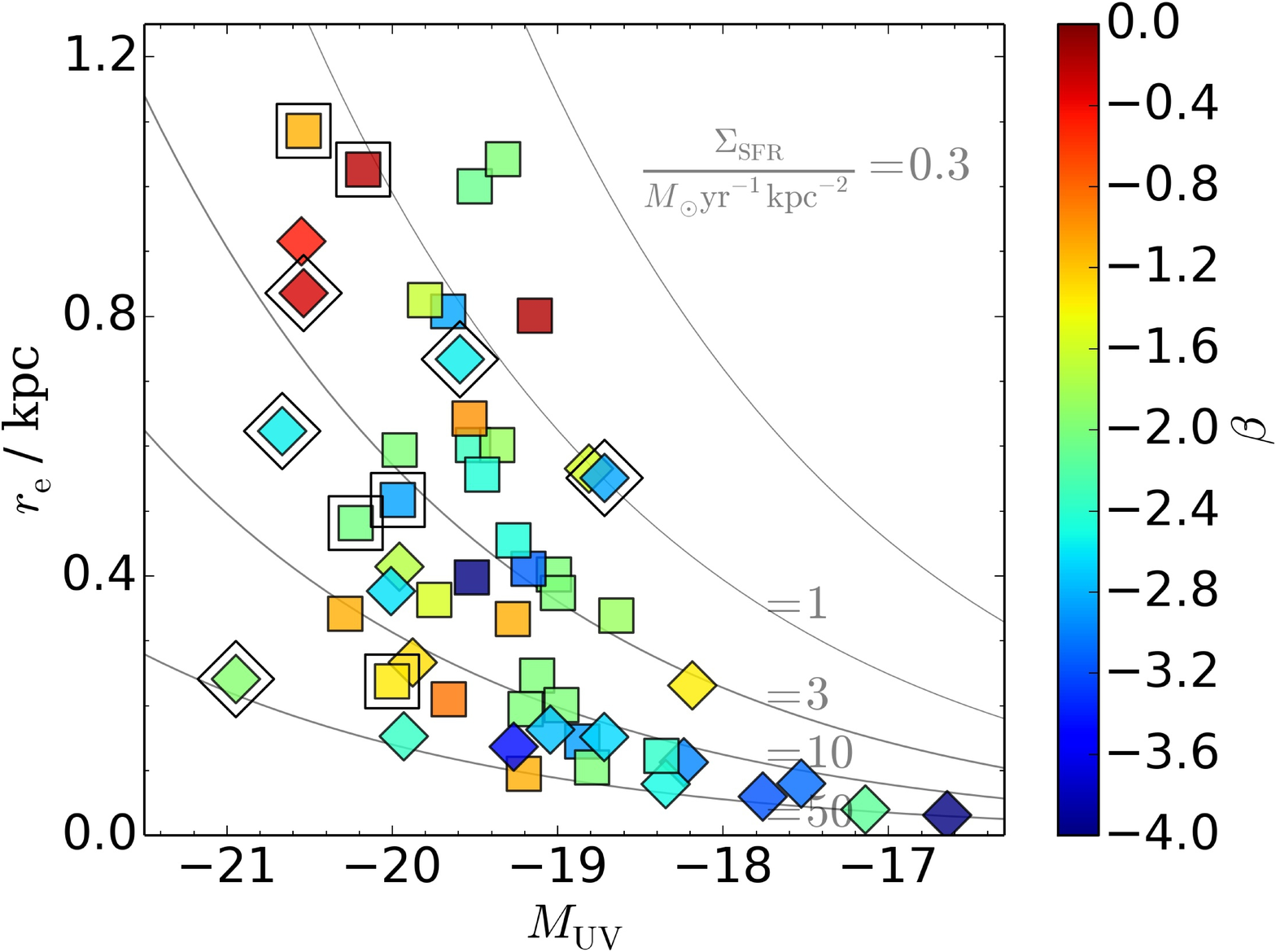}
  \caption{Size--luminosity relation for the merged sample of 
              $z \sim 6-8$. 
              Rhombuses represent galaxies from the cluster field 
              and squares from the parallel and blank fields.
          Galaxies are color-coded by the UV power-law index, 
              $\beta$.
          The solid lines correspond to constant 
              star formation surface densities of
      $\Sigma_\mathrm{SFR} / (M_{\odot} \mathrm{yr^{-1}} \mathrm{kpc}^{-2}) 
              = 0.3, 1, 3, 10, 50$.
          Galaxies with multiple cores are marked by 
              a large, open rhombus or square.}
  \label{fig:magrad_wbeta}
\end{figure}

Figure \ref{fig:magrad} shows the distribution of our galaxies 
in the $r_{\mathrm{e}}$--$M_{\mathrm{UV}}$ plane 
for the two redshift ranges. 
Also plotted are the galaxies selected in \citet{ono13} 
who investigated the size distribution of HUDF12 galaxies 
over similar redshift ranges using \texttt{GALFIT}.
Because of the difference in the dropout band ($i_{814}$ vs $z_{850}$),
the redshift distribution of our $z\sim 6-7$ galaxies is somewhat
different from that of the HUDF12 galaxies which are distributed
around $z \approx 7$ (Figure \ref{fig:redshift-distribution}).
However, since there seems to be little size evolution between 
$z\sim 6$ and $z\sim 7$ (see Figure \ref{fig:radiusz}), we merge these two 
samples into one to improve the statistics.

The merged sample is less affected by cosmic variance 
than the HUDF12 sample. Calculation based on \citet{robertson14} finds 
that at $z\sim 7$ the cosmic variance in the number density of 
$M_{\mathrm UV} = -20$ galaxies reduces from $50\%$ from $30\%$ by adding our 
sample to the HUDF12 one. Similarly, a reduction from $60\%$ to 
$40\%$ is expected at $z\sim 8$.

Similar to \citet{ono13}, we find a positive correlation between 
half-light radius and luminosity in our sample for both redshift 
ranges. 
The average size of our galaxies as a function of luminosity 
is also in a rough agreement with the result of \citet{ono13}, 
once the statistical uncertainties are taken into account.
However, at both redshift ranges,
the correlation seen in our sample is much weaker than
that of \citet{ono13}.
For example, around $M_\mathrm{UV} \sim -19$ to $-20$, the sizes
of our $z\sim 6-7$ galaxies are distributed over 0.1 kpc -- 1.0 kpc.
This is perhaps a consequence of the much larger sample size and
less cosmic variance of our sample as compared with the sample of
\citet{ono13}.

\citet{huang13} have also found a large scatter in size 
for dropout galaxies at $z \sim 4$ and $5$. 
A relatively weak correlation between size and luminosity 
may be common at high redshifts.

Our sample contains two very bright and compact galaxies, 
C-Y1 at $z\sim 8$ and C-i2 at $z\sim 6-7$. 
Both galaxies have also been identified in 
previous papers analyzing the HFF data.
C-Y1 has been reported by \citet{laporte14} to have 
$r_\mathrm{e} = 0.35 \pm0.15\ \mathrm{kpc}$ and 
$M_{\mathrm UV} = -20.88 \pm 0.04$, consistent with our measurements. 
C-i2 has been identified by \citet{atek14} with 
$M_{\mathrm UV} = -20.45 \pm 0.03$, also consistent with our value.

There is no faint and large galaxy found in our sample. 
At $z\sim8$, it is unlikely that we are missing a significant 
fraction of such galaxies due to selection bias, 
because our galaxies are distributed in the 
$r_{\mathrm{e}}$--$M_{\mathrm{UV}}$ plane 
well separate from the boundary of 
$50\%$ detection completeness shown 
by the dashed and dash-dot lines.
However, since some of our galaxies at $z\sim 6-7$ 
are close to the completeness lines, 
their size distribution can be affected 
by the completeness effect.
These completeness lines are determined by the following process.
(1) We generate model galaxies over wide magnitude and size ranges 
and put them randomly on the detection image.
(2) We run \texttt{SExtractor} on this image in exactly the same manner as 
for the original image. (3) The detection rate is found to 
decrease with increasing size. The size at which the detection 
rate is $50\%$ is obtained as a function of absolute magnitude.

The blue points with error bars in each panel of 
Figure \ref{fig:magrad} indicate the average size and the scatter
in the given luminosity bin for the merged sample 
of this study and \citet{ono13}.
The average correlations at two redshifts agree well 
with each other, indicating that the size--luminosity relation
does not significantly evolve from $z\sim 6-7$ to $\sim 8$.
Therefore, we combine the two redshift ranges together to 
examine the dependence of the size--luminosity relation 
on two other galaxy properties, UV color and multiplicity.

\vspace{5pt}
{\it UV color.} 
Figure \ref{fig:magrad_wbeta} plots half-light radii against 
UV luminosities for all the galaxies in the combined sample, 
colored according to the UV slope $\beta$ that is 
defined as $f_{\lambda} \propto \lambda^{\beta}$ with $f_{\lambda}$ 
being the UV continuum flux density with respect to wavelength 
$\lambda$.
We calculate $\beta$ using the equations given in \citet{bouw13}:
$\beta = -2.0 + 4.39(J_{125} - H_{160})$ for $z\sim 6-7$ and 
$\beta = -2.0 + 8.98(JH_{140} - H_{160})$ for $z\sim 8$.
For each redshift, the range of $\beta$ in our sample is consistent 
with those of \citet{bouw13} and \citet{dunlop13}.

We find that largest ($> 0.8$ kpc) galaxies tend 
to be bright ($M_\mathrm{UV} \lesssim -19.5$)
while the remaining galaxies have a wide range of luminosity 
with a weak correlation with size,
as already seen in each panel of Figure \ref{fig:magrad} 
with smaller statistics. 
We also find that largest ($> 0.8$ kpc) galaxies are 
mostly red and smallest ($<  0.08$ kpc) 
galaxies are mostly blue, while the remainings do not show 
a very strong trend. 
There are mainly three factors that make galaxies red: 
high dust extinction, old age, and high metallicity.
Since these three characteristics are often seen in 
evolved galaxies, selecting largest galaxies may lead to
effectively picking out evolved galaxies 
at the redshifts studied here.

We note that the faintest galaxy in the sample, C-i10, 
with $M_\mathrm{UV}=-16.6$, is also the bluest ($\beta=-4.00$)
and smallest ($r_\mathrm{e} = 0.03\ \mathrm{kpc}$) object.
This interesting galaxy, detected thanks to the high lensing magnification of $\mu\simeq 11$,
may be a very young galaxy with an extremely low metallicity.
This galaxy has also been reported by \citet{atek14b} to have 
$M_{\mathrm UV} = -15.80 \pm 0.16$.

\vspace{5pt}
{\it Multiplicity.} 
We also examine if galaxies with multiple cores have any 
preference in size or luminosity.
Multiple cores can be regarded as a sign of a recent merging event.
Many papers have estimated the fraction of galaxies with multiple cores
at high redshift 
\citep[e.g.][]{ravindranath06, lotz08,  
oesch10, law12,guo12, jiang13b}.
For example, \citet{ravindranath06} have reported that 30\% of $z\sim 3$ 
LBGs have multiple cores,
and \citet{jiang13b} have found that $40\%-50\%$ of bright 
($M_\mathrm{UV} \leq -20.5$) galaxies at $5.7\leq z \leq 7.0$ 
have multiple cores.

We identify galaxies with multiple cores in our sample by visual inspection, 
considering the claim by \citet{jiang13b} that while 
galaxies at $z\gtrsim6$ are too small and faint for 
quantitative morphological analysis, visual inspection is 
still valid for examining whether or not a galaxy has 
multiple cores.
Galaxies with multiple cores are marked with a large square 
in Figure \ref{fig:magrad_wbeta}, and marked with a star 
in Figures \ref{fig:fitresults7_1}--\ref{fig:fitresults8_parallel}.
Ten galaxies, or $19\%$ of the sample, are found to have 
multiple cores.
This fraction is similar to that derived by \citet{oesch10} 
at similar redshifts.
As seen in the galaxy images summarized in Appendix,
for most of the galaxies with multiple cores,
the primary cores are distinct compared to the secondary or later cores,
which perhaps implies relatively minor mergers.

As can be seen from Figure \ref{fig:magrad_wbeta}, most of 
the galaxies with multiple cores are bright  
($M_\mathrm{UV} \lesssim -20$), qualitatively consistent with 
the trend seen in the sample of \citet{oesch10}
that brighter galaxies tend to have multiple cores.
More specifically, in the sample of \citet{oesch10},
the brightest and fourth brightest galaxies have 
multiple cores among the 16 $z\sim7$ galaxies.
In our sample, three of the four brightest galaxies ($M_\mathrm{UV} \leq -20.5$) 
at $z\sim 8$ have multiple cores.
On the other hand, we find that the sizes of galaxies 
with multiple cores are distributed widely from 0.2 kpc to 1 kpc.

At $z\sim 6-7$, the bright galaxies 
($-20 \leq M_\mathrm{UV} \leq -19.0$) from the cluster field are on average
smaller than those from the parallel field.
We find about a factor of two difference in 
the average size of those galaxies ($0.32\pm 0.23$ kpc for 
the cluster field and $0.67\pm 0.20$ kpc for the parallel field). 
This discrepancy might be caused by sample variance and/or 
cosmic variance.

\subsection{Redshift evolution of size}

\begin{figure}[tbp]
  \centering
      \includegraphics[width=\linewidth]{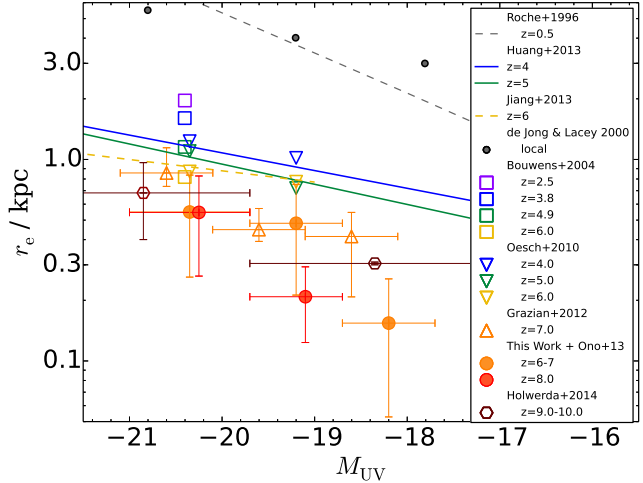}
  \caption{Size--luminosity relations for $2.5 \leq z \lesssim 9-10$ 
              LBGs, overplotted with those for local spiral 
              galaxies and $z\sim 0.5$ irregular galaxies. 
              Our samples combined with \citet{ono13}'s are shown 
              by orange ($z\sim 6-7$) and red ($z\sim 8$) 
              filled circles.
              The purple, blue, green, and yellow open squares 
              are for $z\sim2.5$, $z\sim3.8$, $z\sim4.9$, 
              and $z\sim6$ LBGs by \citet{bouw04}; 
              the blue, green, and yellow open inverse triangles 
              for $z\sim4$, $z\sim5$, and $z\sim6$ LBGs by 
              \citet{oesch10}; 
              the orange triangles for $z\sim7$ LBGs 
              by \citet{graz12};
              and the brown open hexagons for $z\sim9-10$ LBGs 
              by \citet{holwerda14}.
              The blue and green lines show the average relations 
              for LBGs at $z\sim 4$ and $z\sim 5$ by \citet{huang13}, 
              and the yellow dashed line the average relation 
              for Ly$\alpha$ emitters and LBGs at $z\sim 5.7-6.5$ 
              by \citet{jiang13b}. 
              The black dots represent the average relation for 
              local spiral galaxies by \citet{dejonglacey00} 
              and the black dotted line is for $z\sim 0.5$ 
              irregular galaxies by \citet{roche96}.
              The error bars in $r_\mathrm{e}$ are the 1$\sigma$ standard 
              deviations while those in $M_\mathrm{UV}$ correspond to
              the bin widths.}
  \label{fig:grazian}
\end{figure}

Figure \ref{fig:grazian} shows
the average half-light radius as a function of UV luminosity
for $2.5 \leq z \lesssim 9-10$ LBGs 
and for $z\sim 0$ spirals and $z\sim 0.5$ irregulars for comparison. 
Galaxies from our merged sample are plotted as orange 
($z\sim 6-7$) and red ($z\sim 8$) filled circles.
\citet{huang13}, \citet{jiang13b}, \citet{ono13}, 
and \citet{holwerda14} have used \texttt{GALFIT} to measure sizes 
and luminosities, while \citet{bouw04} have used half-light radii 
based on Kron-style magnitudes, and \citet{oesch10} and \citet{graz12} 
based on \texttt{SExtractor}.

From this figure, we find that the average size around $M_\mathrm{UV} = -20.4$ 
gradually becomes smaller with redshift from $z\sim 2.5$ to 
$z\sim 7$ but the evolution from $z\sim7$ to $z\sim 9-10$ 
is not significant.
The slopes of the size--luminosity relation for 
$z\sim6-8$ galaxies seem to be 
steeper than those for $z\sim4-5$ galaxies, although the statistical uncertainty is still large.
This may indicate that fainter, or less massive, galaxies grow 
in size more rapidly over $z\sim 4-8$.
It is worth noting that among the two local ($z\lesssim 0.5$) 
galaxy populations, 
irregular galaxies have a steep slope similarly to those of $z\sim6-8$ 
galaxies.

\begin{figure}[tbp]
  \centering
      \includegraphics[width=\linewidth]{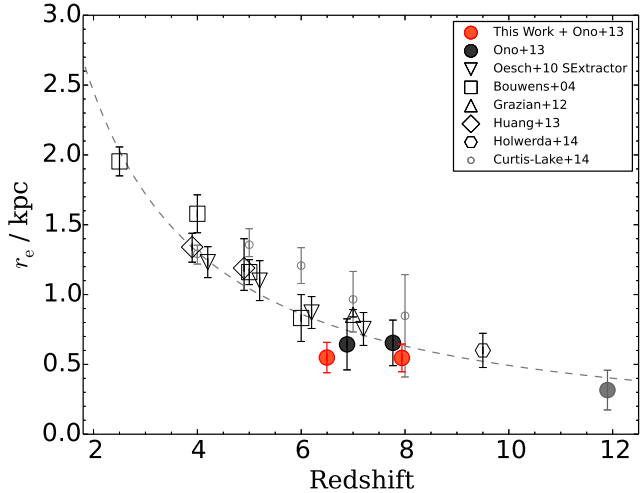}
  \caption{Redshift evolution of the average size of bright 
              galaxies.
          The red circles show the weighted-average radii of our samples 
              combined with \citet{ono13}'s, 
              while the black circles are for \citet{ono13}'s.
          The error bars show the 1$\sigma$ standard error.
}
  \label{fig:radiusz}
\end{figure}

Plotted in Figure \ref{fig:radiusz} is the average half-light 
radius of bright ($(0.3-1) L^{*}_\mathrm{z=3}$) galaxies as 
a function of redshift.
In the calculation of the average radii for the merged samples 
of this work and \citet{ono13}, we reduce the weight of the 
samples from the cluster field according to the uncertainty in 
the mass model.
For the cluster-field sample at each redshift,
the uncertainty in magnification is calculated from the average of 
the variances in magnification among the eight public mass maps at 
the positions of the sample galaxies. 
This uncertainty is then converted into the uncertainty in radius
and is quadratically added to the statistical error in the average 
radius for this sample, thus resulting in a reduction of the weight 
compared with the parallel-field and \citet{ono13}'s samples  
for which only the statistical error is considered.
We include the $z\sim 12$ object given in \citet{ono13}.
The data of \citet{ferg04} are not included because their sample 
includes brighter ($<5L^{*}_\mathrm{z=3}$) galaxies. 
We find that the average size of bright galaxies decreases 
from $z \sim 2.5$ to $6-7$, in agreement with previous results 
\citep[e.g.][]{bouw04, hathi08a, oesch10, graz12, huang13, ono13}, 
while the evolution between $z\sim 6-7$ and $\sim8$ is 
insignificant.
We note that the average radius of our $z\sim6-7$ combined 
sample may be underestimated because all but one are 
in the range of $(0.3-0.5) L^{*}_\mathrm{z=3}$.

The small open circles in Fig. \ref{fig:radiusz} indicate 
the mode of the log-normal distribution of $r_{\mathrm{e}}$ for 
$z \sim 4-8$ LBGs over the same luminosity range obtained by 
\citet{curtislake14}, who have adopted a non-parametric, 
curve-of-growth method to measure sizes.
Their values are slightly but systematically higher than 
the other measurements except for $z\sim 4$, 
leading them to conclude that the typical galaxy size does not
significantly evolve over $z \sim 4-8$.
The reason for this systematic difference is not clear,
although we find in our $z\sim 6-7$ sample that adopting the mode 
instead of the average results in 0.15 kpc {\it decrease}.
Our $z\sim 8$ sample is too small to calculate a modal value.

Fitting the size evolution of $r_\mathrm{e} \propto (1+z)^{-m}$ 
to the data except those of \citet{curtislake14} who adopted 
the modal values 
gives $m=1.24\pm0.1$, which is consistent with previous 
results based on average $r_\mathrm{e}$ measurements 
\citep{oesch10, ono13}.
Analytic models of dark-halo evolution predict that the virial 
radius scales with redshift as $(1+z)^{-1}$ for halos with 
a fixed mass and as $(1+z)^{-1.5}$ for halos with a fixed 
circular velocity \citep[e.g.][]{ferg04}.
The value we find, $m=1.24$, is in the middle of these two cases.
However, any previous attempts to link galaxies 
to dark matter halos using an observed redshift scaling of 
half-light radius have implicitly made a non-trivial assumption
that half-light radius linearly scales with virial radius.

\begin{figure}[tbp]
  \centering
      \includegraphics[width=\linewidth]{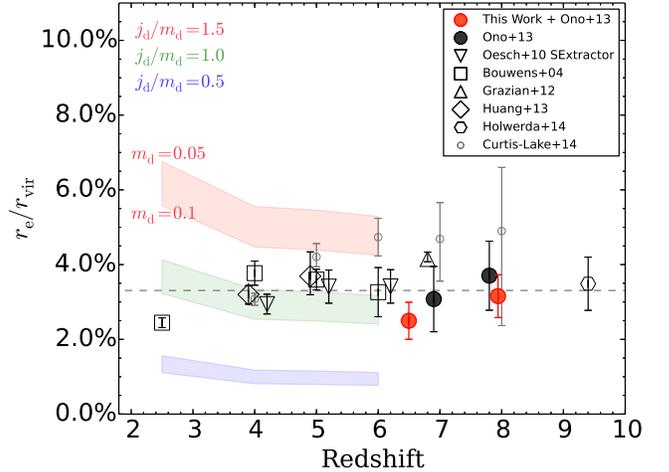}
  \caption{Redshift evolution of the half-light radius 
              to virial radius ratio.
              The error bars show the 1$\sigma$ standard error of the half-light radius.
              The shaded bands are predictions from the model by \citet{mmw98}
              changing the $j_\mathrm{d}/m_\mathrm{d}$ and $m_{\mathrm{d}}$
              within the range of $0.05-0.1$.
              The red, green, and blue bands correspond to 
	       $j_\mathrm{d}/m_\mathrm{d} = 1.5, 1.0, 0.5$. 
	       The upper edge of each band corresponds to $m_\mathrm{d}=0.05$
	       and the lower to $m_\mathrm{d}=0.1$.
	       }
  \label{fig:ratio}
\end{figure}

In order to obtain further insights into disk evolution 
in dark matter halos, we take a different approach.
We combine the so-called abundance matching analysis that connects the stellar 
mass and halo mass with the observed relation between stellar mass and luminosity.
Specifically, we adopt the abundance matching result of \citet{behroozi13}, 
and the observed stellar mass--luminosity relations of 
\citet{reddysteidel09} for $z\sim2.5$ galaxies and 
\citet{gonz11} for $z\sim4-7$ galaxies
\footnote{They showed the relation for $z\sim4$ and state that 
it is consistent with no evolution at $z\sim4-7$} 
to calculate the halo mass of galaxies in Figure \ref{fig:radiusz} from 
their UV luminosity. 
The estimated halo mass of $M_{\mathrm{UV}} = - 20.2$ galaxies 
at $z=6.0$ is $\log(M_{h} / M_{\odot}) = {11.2}$, 
in good accordance with the recent clustering result, 
$\log(M_{h} / M_{\odot}) = {11.0^{+0.4}_{-0.6}}$, by \citet{BN14}.
Then, the virial radius is calculated by
\begin{eqnarray}
r_\mathrm{vir} = \left(\frac{2GM_\mathrm{vir}}{\Delta_\mathrm{vir} \Omega_\mathrm{m}(z) {H(z)}^{2}}\right)^{1/3},
\end{eqnarray}
where $\Delta_\mathrm{vir} \approx (18\pi^{2} + 60 x -32x^{2})/\Omega_\mathrm{m}(z)$ and $x=\Omega_\mathrm{m}(z) - 1$ \citep{bryannorman98}.
The uncertainties in the estimation of virial radii are not 
considered here.
We exclude the $z\sim 12$ data because 
the analysis result of \citet{behroozi13} 
does not extend to that redshift.
We note that \citet{kravtsov13} have conducted 
a similar analysis for local galaxies, and have
found a linear relation between half-mass radius and 
virial radius over eight orders of magnitude in stellar mass.
Our analysis represents the first analysis of the evolution 
of the relation of the galaxy and halo sizes over 
a wide redshift range based on the abundance matching technique.

Figure \ref{fig:ratio} shows the ratio of half-light radius to 
virial radius for galaxies over $z\sim2.5-9.5$.
When limited to the data of average $r_\mathrm{e}$ measurements,
we find the ratio to be virtually constant at $3.3\pm 0.1\%$
over the entire redshift range.
The modal data of \citet{curtislake14} give systematically higher ratios 
over $z\sim5-8$, perhaps showing a slight decrease toward $z\sim 4$, 
but the differences from $3.3\%$ are mostly within 
the $1-2\sigma$ errors.
Thus, the assumption of a constant $r_\mathrm{e}/r_\mathrm{vir}$ ratio 
appears to be broadly consistent with the data.
Our analysis shows that the halo mass of $(0.3-1) L^{*}_\mathrm{z=3}$ 
galaxies mildly decreases with redshift.
This, combined with the constant $r_\mathrm{e}/r_\mathrm{vir}$ ratio
found here, results in the redshift evolution of $1<m<1.5$.

According to the disk formation model by \citet{mmw98}, 
$r_\mathrm{e}/r_\mathrm{vir}$ is described as:
\begin{eqnarray}
\frac{r_\mathrm{e}}{r_\mathrm{vir}} 
&=& 
\frac{1.678}{\sqrt{2}}\left(\frac{j_\mathrm{d}}{m_\mathrm{d}} 
\lambda\right) {f_\mathrm{c}}^{-1/2} f_\mathrm{R}\label{eq:ratio}\\
&\equiv& 
\frac{1.678}{\sqrt{2}} \lambda', 
\end{eqnarray}
for a self-gravitating disk embedded in a NFW dark matter halo, 
where $r_{\mathrm e} = 1.678 R_{\mathrm d}$ 
($R_\mathrm{d}$ is the scale length of the exponential disk), 
$m_\mathrm{d}$ and $j_\mathrm{d}$ are, respectively, the fractions of 
the mass and angular momentum within the halo belonging to the disk,
$\lambda$ the spin parameter of the halo,
$f_\mathrm{c}$ a function of the halo concentration $c_{\mathrm{vir}}$, 
and $f_\mathrm{R}$ a function concerning baryonic contraction.
In the context of this model, our ratio (for average $r_\mathrm{e}$)
gives $\lambda' = 3.3\% \times \sqrt{2}/1.678 \simeq 0.028$.

Although $\lambda$ and $c_{\mathrm{vir}}$ 
are well determined by N-body simulations 
\citep{vitvitska02, davisnatarajan09, prada12}, 
$j_{\mathrm{d}}$ and $m_{\mathrm{d}}$ are not reliably predicted
because complicated baryonic precesses have to be considered.
Since the size ratio, that is $\lambda'$, 
is proportional to $j_\mathrm{d}/m_\mathrm{d}$
while, with a fixed $j_\mathrm{d}/m_\mathrm{d}$, 
being dependent more weakly on $m_\mathrm{d}$
through $f_\mathrm{R}$ and essentially insensitive to $j_\mathrm{d}$,
we here aim to find out 
what value of $j_\mathrm{d}/m_\mathrm{d}$ is reasonable.
The shaded bands in Figure \ref{fig:ratio} are the ratios 
calculated from Equation (\ref{eq:ratio}) 
using the simulation results of $\lambda$ 
by \citet{vitvitska02} and \citet{davisnatarajan09} 
and $c_{\mathrm{vir}}$ by \citet{prada12}
for three $j_\mathrm{d}/m_\mathrm{d}$ values 
over a conservative $m_\mathrm{d}$ range of
$0.05 \le m_\mathrm{d} \le 0.1$.
We find that the observed size ratio is consistent
with the model when $j_\mathrm{d} \sim m_\mathrm{d}$ is assumed.

The ratio of $3.3 \%$ is larger than that for local galaxies
reported by \citet{kravtsov13}, $1.5\%$, implying that the ratio 
decreases from $z\sim 2.5$ to the present day.
Indeed, \citet{kravtsov13} have compared the local value 
with a theoretically plausible value for galaxies 
at the era of disk formation, $\simeq 3.2\%$, 
and attributed the lower local ratio to 
pseudo growth of dark matter halos since disk formation.
In any case, our finding of a constant $r_\mathrm{e}/r_\mathrm{vir}$ ratio 
of $3.3\%$ over a wide redshift range of 
$2.5 \lesssim z \lesssim 9.5$ strongly constrains 
disk formation models.

\subsection{Star-formation Rate Surface Density}

The physical state of star formation of a galaxy is effectively 
described by the total $S\!F\!R$ and 
the $\Sigma_\mathrm{SFR}$.  
While the former is just the scale of star formation,
the latter corresponds to the intensity of star formation 
and is useful for discussing the mode of star formation.
We calculate the $S\!F\!R$ and $\Sigma_\mathrm{SFR}$ for our galaxies with equations (3) 
from \citet{kennicutt98} and (4) from \citet{ono13}, respectively: 
\begin{eqnarray}
\frac{S\!F\!R}{M_{\odot} \rm{yr^{-1}}} 
&=& 1.4 \times 10^{-28} \frac{L_{\nu}}{ \rm{erg \ s^{-1} \ Hz^{-1}}}\\
\Sigma_\mathrm{SFR} 
&=& \frac{S\!F\!R / 2}{\pi {r_\mathrm{e}}^{2}}.
\end{eqnarray}
The weighted-log-average $\Sigma_\mathrm{SFR}$s of $(0.3-1) L^{*}_\mathrm{z=3}$ galaxies 
considering the uncertainty in the mass model are 
$4.1\ M_{\odot} \mathrm{yr}^{-1} \mathrm{kpc}^{-2}$ at $z\sim 6-7$ 
and $5.6\ M_{\odot} \mathrm{yr}^{-1} \mathrm{kpc}^{-2}$ at $z\sim8$,
slightly higher than $3.5\ M_{\odot} \mathrm{yr}^{-1} \mathrm{kpc}^{-2}$ at $z\sim 7$ 
and $3.2\ M_{\odot} \mathrm{yr}^{-1} \mathrm{kpc}^{-2}$ at $z\sim8$ by \citet{ono13}.

As found from Figure \ref{fig:magrad_wbeta}, 
our galaxies are forming stars with a rate of $S\!F\!R \sim 1-10\ M_{\odot} \mathrm{yr}^{-1}$ 
and with a surface intensity of $\Sigma_\mathrm{SFR} \sim 1-50\ M_{\odot} \mathrm{yr}^{-1} \mathrm{kpc}^{-2}$.
This $\Sigma_\mathrm{SFR}$ range is slightly wider toward higher values than 
reported by \citet{ono13} based on HUDF12, 
reflecting the fact that 
our galaxies are distributed over a wider area 
in the size--luminosity plane than those of \citet{ono13}.
Our sample extends especially toward smaller half-light radii.

\begin{figure}[tbp]
  \centering
      \includegraphics[width=\linewidth]{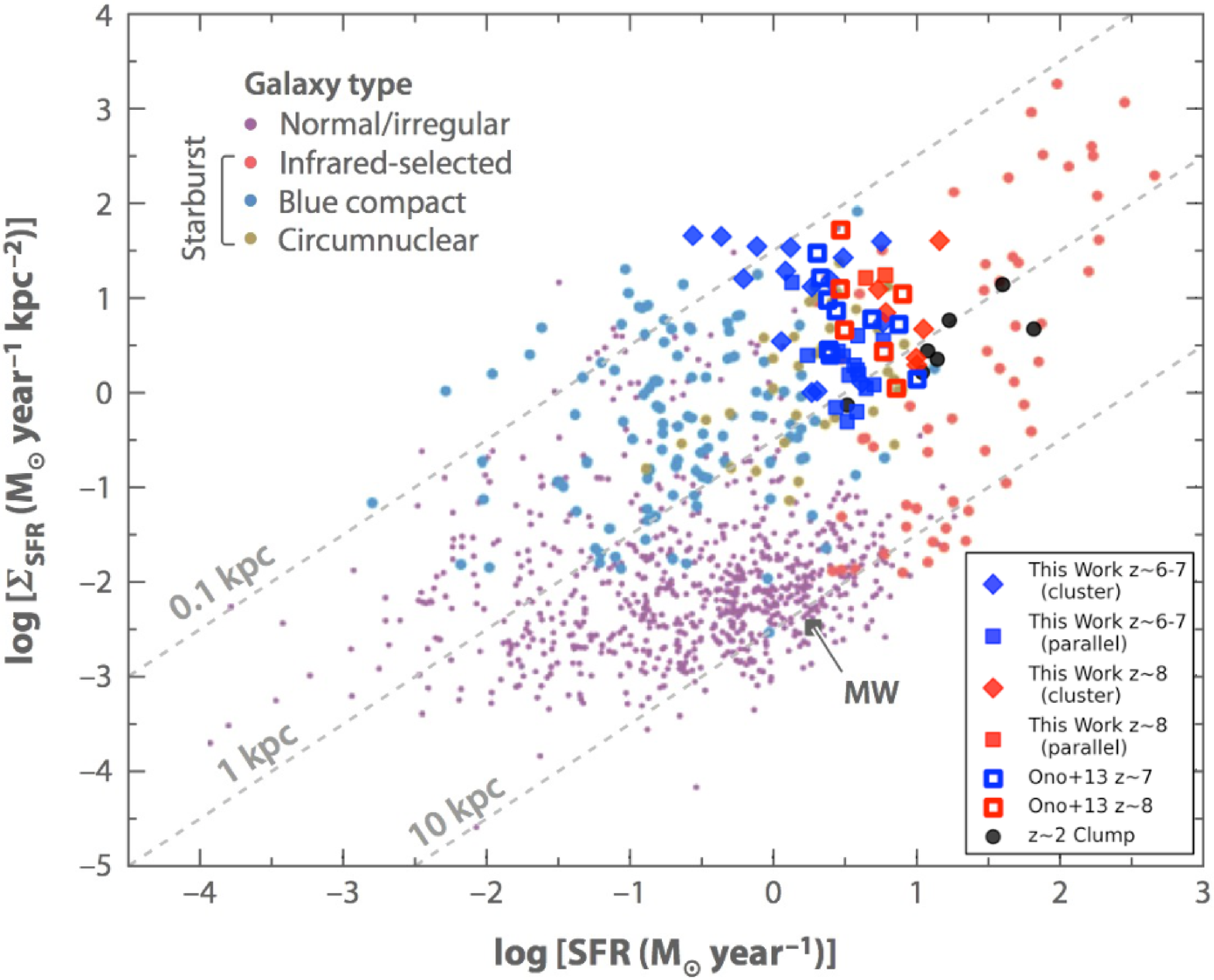}
  \caption{Distribution of our galaxies in the $S\!F\!R$--$\Sigma_\mathrm{SFR}$ plane,
              overplotted with a various types of local galaxies 
              and star-forming clumps of $z\sim2$ galaxies \citep{genzel11}.
              Plotted on Fig. 9 in \citet{keneva12}.}
  \label{fig:SFRsigmaSFR}
\end{figure}

We show in Figure \ref{fig:SFRsigmaSFR} the distribution 
in the $S\!F\!R$--$\Sigma_\mathrm{SFR}$ plane of our galaxies and various types of
local galaxies, in order to examine in what sense the state 
of star formation of $z\sim 6-8$ galaxies are similar or 
dissimilar to local ones.
Comparison to normal galaxies finds that our galaxies have 
much higher (typically three orders of magnitude higher) 
$\Sigma_\mathrm{SFR}$s than normal galaxies in spite of having modest $S\!F\!R$s 
similar to that of the Milky Way.
In other words, $z\sim 6-8$ galaxies are forming stars 
at similar rates to local normal galaxies 
but in $10^3$ times smaller areas.

Our galaxies are roughly comparable in $\Sigma_\mathrm{SFR}$ to average 
infrared-selected galaxies and to blue compact galaxies,
while falling between these two galaxy populations in $S\!F\!R$.
It is circumnuclear regions that resembles our galaxies most.
Circumnuclear regions are not the whole bodies of galaxies 
but starbursting rings at the center of a certain type of 
galaxies in which gas is effectively fed along bars.
This resemblance may suggest that $z \sim 6-8$ galaxies 
have a similar amount of cold gas of a similarly high density 
to that of circumnuclear regions.

\citet{ono13} have found that the $\Sigma_\mathrm{SFR}$s of $z \sim 7-8$ galaxies 
are comparable to those of infrared-selected galaxies and 
circumnuclear regions.
While we confirm this finding above, we also find that 
infrared-selected galaxies are scaled-up systems 
in terms of $S\!F\!R$.

Finally, we find that our galaxies have similar $\Sigma_\mathrm{SFR}$s but 
slightly lower $S\!F\!R$s than star-forming clumps of $z\sim 2$ 
galaxies taken from \citet{genzel11} (black dots in Figure \ref{fig:SFRsigmaSFR}).
In this sense, $z\sim 6-8$ galaxies may be considered to be 
scaled-down systems of clumps.

In this subsection we have neglected dust extinction.
If we use UV slope to calculate the extinction at 1600\AA,
$A_{1600}$, according to \citet{meurer99}'s formula, 
we find a median $A_{1600} = 1.2$ in our sample, 
corresponding to a factor 2.9 increase in $S\!F\!R$ (and $\Sigma_\mathrm{SFR}$).
Therefore, if \citet{meurer99}'s formula is still applicable 
to $z\sim 6-8$ galaxies (although they could have very different 
stellar populations and dust properties from local starbursts),
then a significant fraction of our galaxies enter the region 
in the $S\!F\!R$--$\Sigma_\mathrm{SFR}$ plane occupied by local infrared-selected 
galaxies.

\section{Conclusions}
We have used the complete data of the Abell 2744 cluster and parallel fields 
taken in the HFF program  
to measure the intrinsic size and magnitude of 
31 $z\sim6-7$ galaxies and eight $z\sim 8$ galaxies
by directly fitting light profiles of observed galaxies with lensing-distorted 
S\'ersic profiles on the image plane with the \texttt{glafic} software.
The lensing effect has been calculated using the mass map  
constructed by \citet{ishigaki15}.
Our sample includes a very faint galaxy with $M_\mathrm{UV} = -16.6$
that is detected thanks to high magnification.
Combining with the HUDF12 sample of \citet{ono13} that is based on essentially 
the same method of two-dimensional S\'ersic profile fitting,
we now have uniform samples of 40 $z\sim6-7$ galaxies and 
14 at $z\sim8$ galaxies with accurate size measurements.
These large samples enable us to study the statistics of
galaxy sizes and their dependence on other physical parameters
at these high redshifts.
The followings are the main results obtained in this paper.
\\

(i) We have found that a correlation between the half-light radius
and UV luminosity indeed exists, but it is not very tight.
We have also found that largest ($r_\mathrm{e}>0.8$ kpc) galaxies are 
mostly bright and red in UV color while smallest ($r_\mathrm{e}<0.08$ kpc) 
ones mostly blue, and that galaxies with multiple cores tend to 
be bright.
\\

(ii) We have compared the size--luminosity relation 
at $z\sim6-8$ with those of LBGs over $2.5 \lesssim z \lesssim 9-10$
taken from the literature. 
The average size of bright galaxies decreases from $z \sim 2.5$ to 
$z\sim 7$ but the evolution slows down beyond $z \sim 7$.
Irregular galaxies at $z\sim0.5$ have 
a similarly steep slope of the size--luminosity relation 
to $z\sim6-8$ LBGs.
The average size of bright ($(0.3-1)L^\star_{z=3}$) 
galaxies scales as $(1+z)^{-m}$ with $m=1.24\pm0.1$
over $2.5 \lesssim z \lesssim 12$, which is consistent 
with previous studies \citep{oesch10, ono13}.
\\

(iii) 
We have used the abundance matching results by \citet{behroozi13}
to find that the ratio of half-light radius to virial radius is
virtually constant at $3.3\pm 0.1 \%$ over $2.5 \lesssim z \lesssim 9.5$.
This constant ratio is in good agreement with 
the disk formation model by \citet{mmw98}
with plausible values of parameters describing the halo structure, 
if we take $j_\mathrm{d} \sim m_\mathrm{d}$.
\\

(iv) The $\Sigma_\mathrm{SFR}$s of $z\sim6-8$ galaxies are typically 
three orders of magnitude higher than those of local normal 
spiral galaxies.
The distribution of our galaxies in the $S\!F\!R$--$\Sigma_\mathrm{SFR}$ plane is
largely overlapped with that of circumnuclear star-forming 
regions in local barred galaxies, 
which may suggest a similarity in the environment of star formation.
\\

This is a first report of size analysis using the data from the HFF.
The complete observations of the six HFF clusters will provide us with 
more statistically significant samples, 
including very faint galaxies 
that have never been investigated.
Size measurements for these new samples will help 
advance our understanding of galaxy formation and evolution 
through galaxy size studies, which provide 
complementary information to luminosity and color studies.
\\

We would like to thank Yoshiaki Ono for his helpful 
comments and kindly providing with us the HUDF12 data.
This work was supported in part by World Premier International Research Center Initiative (WPI Initiative), MEXT, Japan, and Grant-in-Aid for Scientific Research from the JSPS (26800093).

Facilities: \facility{HST(ACS/WFC3IR)}.

\bibliographystyle{apj}
\bibliography{bibtex.bib}

\newpage

\begin{figure*}

\appendix
\section*{Supplemental Figures}
\label{galaxy_figures}

\begin{center}
      \includegraphics[width=0.6\linewidth]{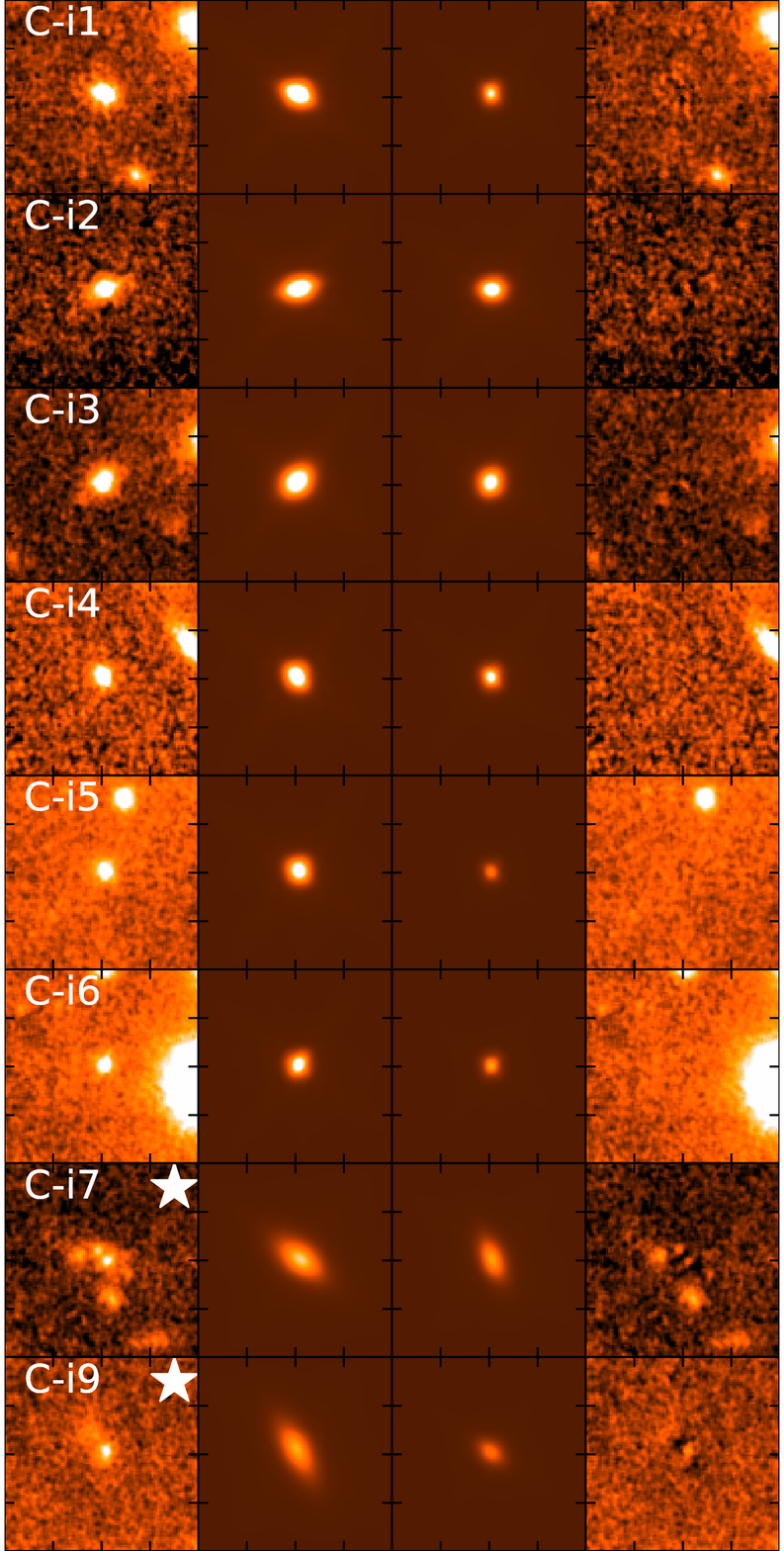}
  \caption{Fitting results for $z\sim6-7$ galaxies 
              in the cluster field. 
              From left to right, $3''\times 3''$ cut-out images, 
              best-fit S\'ersic profiles on image plane, 
              best-fit S\'ersic profiles on source plane, 
              and residual images.
              Galaxies with multiple cores are marked with a star.
              A high resolution version of Figures \ref{fig:fitresults7_1}--\ref{fig:fitresults8_parallel} is available at {\url{http://hikari.astron.s.u-tokyo.ac.jp/\~kawamata/articles/size-abell2744/supplemental-figures}}.}
  \label{fig:fitresults7_1}
\end{center}
\end{figure*}

\begin{figure*}
\begin{center}
      \includegraphics[width=0.6\linewidth]{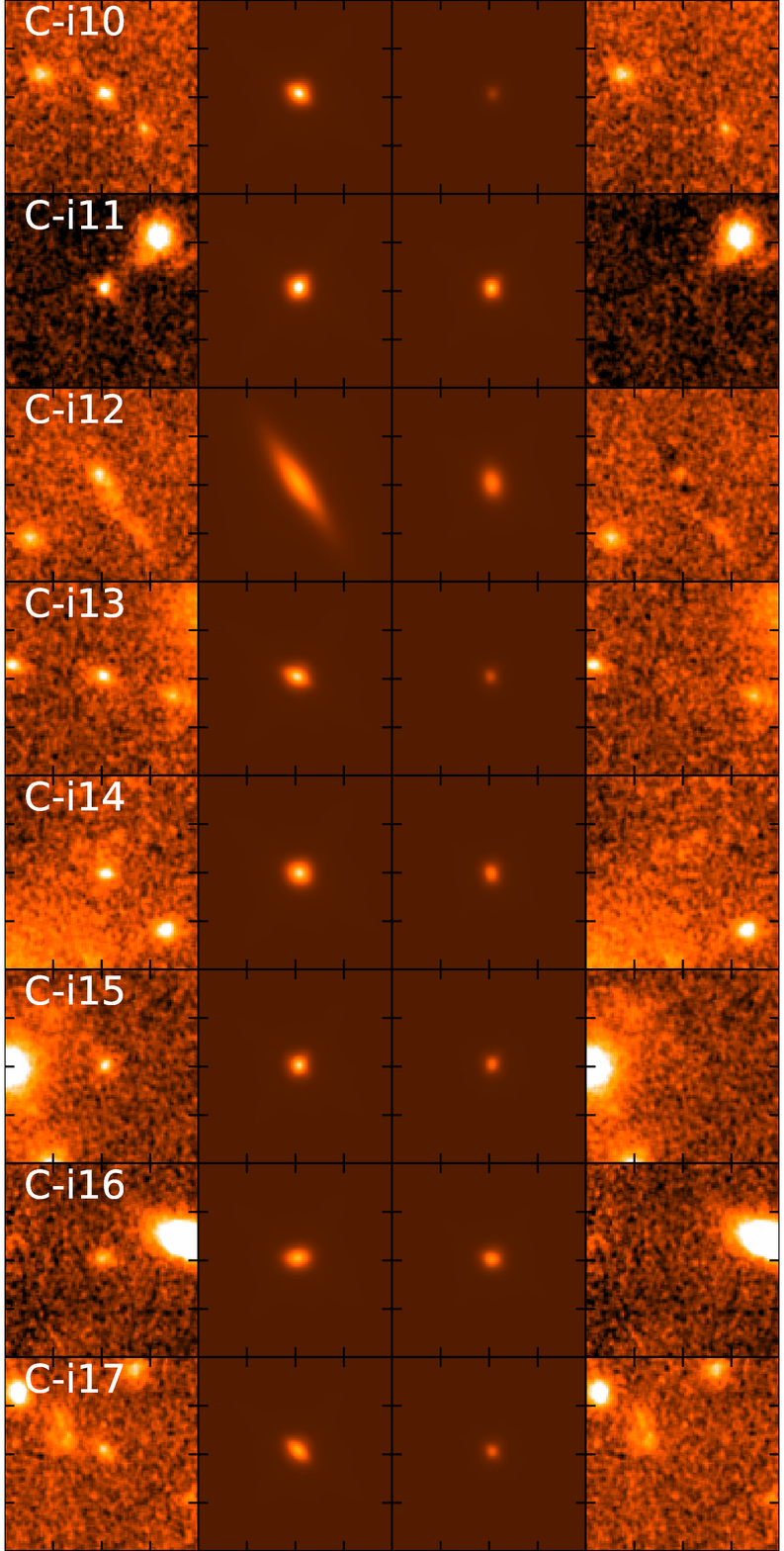}
  \caption{Continuation of Figure \ref{fig:fitresults7_1}}
  \label{fig:fitresults7_2}

\end{center}
\end{figure*}

\begin{figure*}
\begin{center}

      \includegraphics[width=0.6\linewidth]{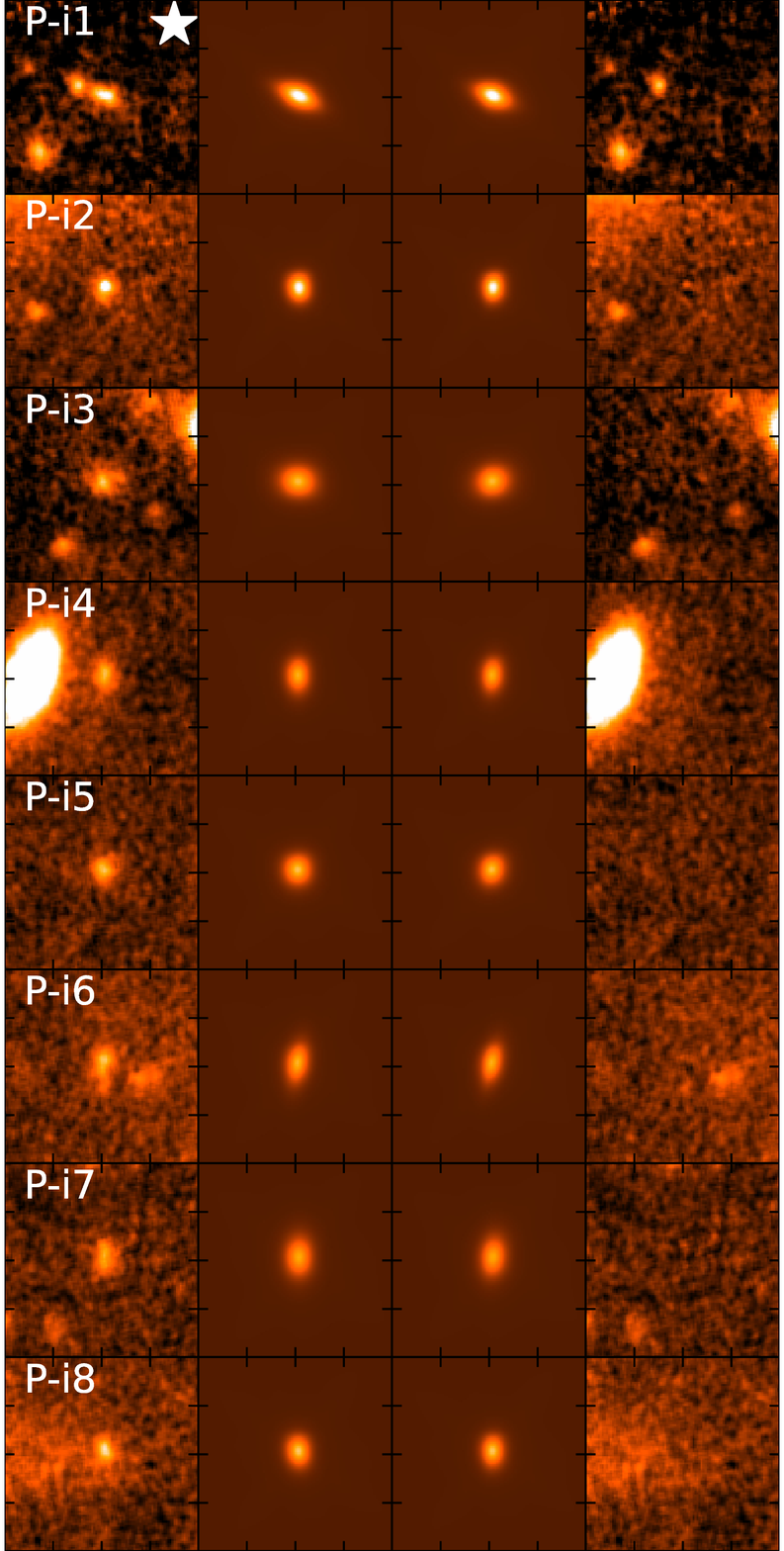}
  \caption{Same as Figure \ref{fig:fitresults7_1} 
              but for $z\sim6-7$ galaxies in the parallel field.}
  \label{fig:fitresults7_1_parallel}
\end{center}
\end{figure*}

\begin{figure*}
\begin{center}
      \includegraphics[width=0.6\linewidth]{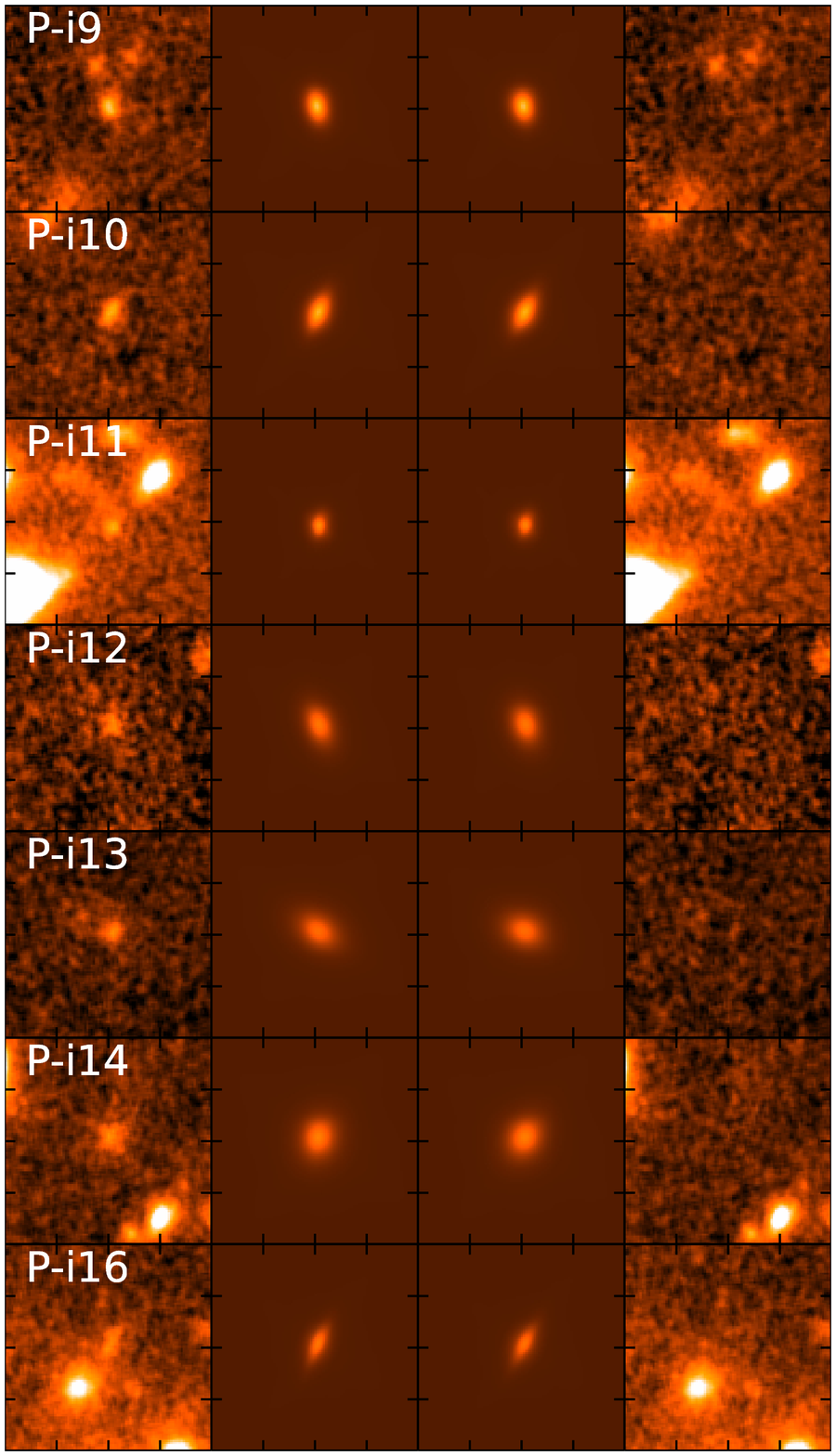}
  \caption{Continuation of Figure \ref{fig:fitresults7_1_parallel}}
  \label{fig:fitresults7_2_parallel}
\end{center}
\end{figure*}

\begin{figure*}
\begin{center}

      \includegraphics[width=0.6\linewidth]{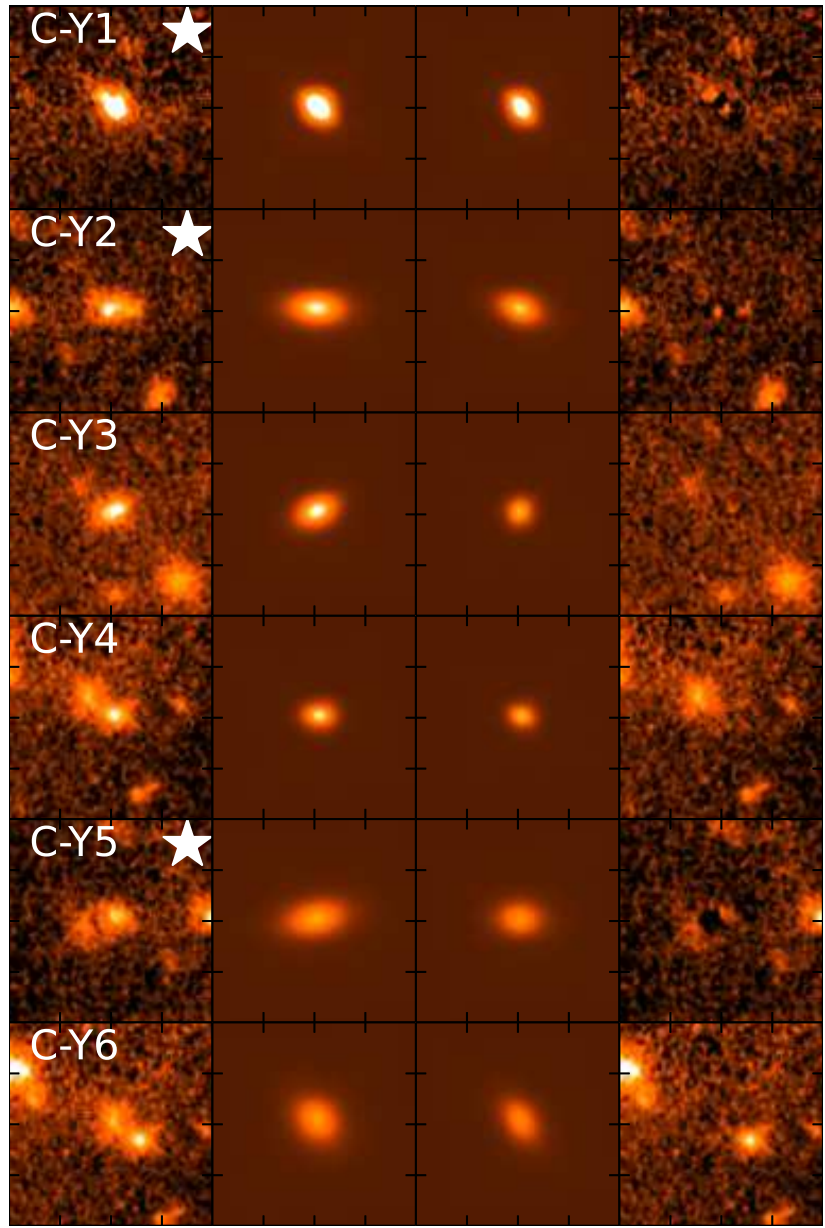}
  \caption{Same as Figure \ref{fig:fitresults7_1} 
              but for $z\sim8$ galaxies in the cluster field.}
  \label{fig:fitresults8}
\end{center}
\end{figure*}

\begin{figure*}
\begin{center}

      \includegraphics[width=0.6\linewidth]{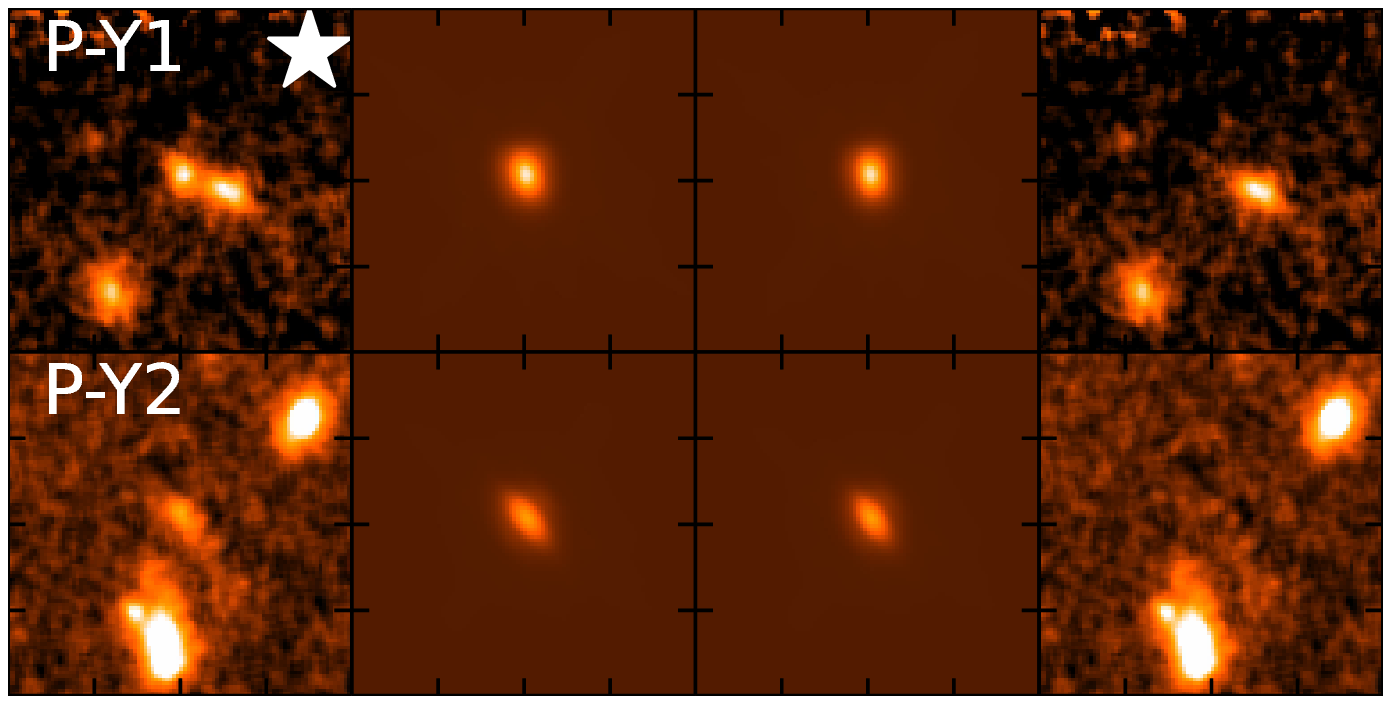}
  \caption{Same as Figure \ref{fig:fitresults7_1} 
              but for $z\sim8$ galaxies in the parallel field.}
  \label{fig:fitresults8_parallel}
\end{center}
\end{figure*}

\end{document}